\documentclass[12pt]{article}
\usepackage{amssymb,graphicx}
\usepackage{epstopdf}
\usepackage{amsmath,amsfonts}
\usepackage{epsfig}

\def\d{\partial}
\def\l{\left(}
\def\r{\right)}

\newcommand{\be}{\begin{equation}}
\newcommand{\ee}{\end{equation}}
\newcommand{\bea}{\begin{eqnarray}}
\newcommand{\eea}{\end{eqnarray}}
\newcommand{\bg}{\begin{gather}}
\newcommand{\eg}{\end{gather}}
\newcommand{\bseq}{\begin{subequations}}
\newcommand{\eseq}{\end{subequations}}

\newcommand{\sfrac}[2]{{\textstyle\frac{#1}{#2}}}
\begin{document}
\begin{flushright}
CERN-PH-TH/2005-265\\
HUTP-05/A0056
\end{flushright}
\vspace{10pt}

\begin{center}
  {\LARGE \bf Null energy condition \\[0.3cm] and superluminal propagation } \\
\vspace{20pt}
%\medskip
S.~Dubovsky$^{a,b}$, T.~Gr\'egoire$^{a,c}$, A.~Nicolis$^d$, R.~Rattazzi$^a$\\
\vspace{15pt}
  $^a$\textit{
Department of Physics, CERN Theory Division, CH-1211 Geneva 23, Switzerland
  }\\
\vspace{5pt}
$^b$\textit{
Institute for Nuclear Research of
         the Russian Academy of Sciences,\\  60th October Anniversary
  Prospect, 7a, 117312 Moscow, Russia}\\
\vspace{5pt}
$^c$\textit{Physics Department, Boston University, \\
590 Commonwealth Ave., Boston, MA 02215, USA}\\
\vspace{5pt}
$^d$\textit{Jefferson Physical Laboratory, Harvard University, Cambridge MA 
02138,
USA} 
    \end{center}
    \vspace{5pt}

\begin{abstract}
We study whether a violation of the null energy condition necessarily
implies the presence of instabilities. We prove that
this is the case in a large class of situations, including isotropic
solids and fluids relevant for cosmology. On the other hand we present
several counter-examples of consistent effective field theories possessing a 
stable background where the null energy condition is violated.
Two necessary features of these counter-examples are the lack of isotropy of the background and
the presence of superluminal modes.  We argue
that many of the properties of massive gravity can be understood by
associating it to a solid at the
edge of violating the null energy condition.  We briefly analyze the difficulties of mimicking 
$\dot H>0$ in scalar tensor theories of gravity.

\end{abstract}
%\newpage \tableofcontents \newpage
\section{Introduction}
Energy conditions play an important role in general relativity and are a basis of singularity theorems \cite{Hawkingtheorems} and entropy bounds \cite{Bousso}.
% They are also  relevant for cosmological applications.  
These conditions reflect the microscopic properties
of the medium sourcing the energy momentum tensor. In some cases  they may be violated without making the system physically unacceptable. This
is the case of the strong energy condition, whose violation in Nature is indeed
suggested by the success of the theory of inflation and also by the recent
observation of cosmic acceleration.
 It is however widely believed that any well behaved physical systems should
  respect the weakest of these conditions stating that $T_{\mu\nu}n^\mu n^\nu\geq 0$ for any null vector $n$. This condition is often referred to as the null energy condition (NEC). For an isotropic fluid with positive energy density, NEC amounts to $w\equiv p/\rho\geq -1$.
  In some simple cases a connection between violation of NEC and the presence of instabilities in the system sourcing $T_{\mu\nu}$ was already established (see for instance \cite{Hsu:2004vr}). It would however be desirable to investigate under what condition this connection becomes a theorem.
%On the other hand, some
%of the conditions which were thought to be solid in the past turned out to 
%be violated in Nature. 
%Thus a spectacular success of the inflationary models in resolving
% the puzzles of the Big Bang scenario and in predicting the 
%spectrum of the primordial fluctuations strongly suggests that the energy
%density of the Universe was dominated by the fluid violating the strong energy
%condition just after the Big Bang. Even more convincing argument favoring the 
%existence of fluids violating the strong energy condition is provided by the
%recent observations indicating the 
%accelerated expansion of the Universe.
%While it is fairly easy to find consistent field theories 
%violating the strong energy condition, violation of the null energy
%condition (NEC) 
%to the best of our knowledge implies the presence of classical or
%quantum instabilities in all known examples (see, e.g. Refs.
%for some examples). This suggests that there
%is a kind of theorem relating violation of NEC and the presence of
%instabilities in the system.  
This question is obviously theoretically interesting\footnote{For
instance, NEC prevents the existence of Lorentzian wormholes~\cite{wormholes} and of
flat bouncing cosmologies.},
but it may also turn out to be relevant to
 present day cosmology. Indeed, recent supernova data
  mildly favor  a NEC-violating 
equation of state~\cite{omega<-1}, $w<-1$. If this result were to be confirmed by  
 more precise experiments in the future, it would become essential to understand
 under what conditions NEC can be violated in the absence of pathologies.
 A no-go theorem valid in ordinary 4D theories may point towards more exotic 
 scenarii, like for instance the Dvali-Gabadadze-Porrati model \cite{Dvali:2000hr}
where an effective $w$ slightly less than $-1$ for dark energy can be 
achieved~\cite{Sahni:2002dx,Lue:2004za},
without manifest  pathologies~\cite{Nik}.
Finally, as we will show, the study of NEC violating field theories
 opens up a new perspective on  theories of massive gravity 
(see Refs.
\cite{Arkani-Hamed:2002sp,Arkani-Hamed:2003uy,Rubakov:2004eb,Dubovsky:2004sg}).

The purpose of the current 
paper is to study in detail the relation between
instabilities and NEC violation. For most  of the paper we will be
dealing with a system of derivatively coupled scalar fields with coordinate
dependent condensates. Although this system may seem somewhat artificial, 
its simplicity allows us to concentrate on the physically relevant features
avoiding technicalities which arise when one includes higher spin
fields (e.g. vectors). Moreover, as we will explain in section~\ref{fluids} our
results are in fact significantly more general. In particular our arguments
apply to arbitrary relativistic solids and fluids. The main reason for this generality is
that NEC violation  is related to instabilities in a particular
sector of the theory. This sector consists of just the Goldstone
degrees of freedom (sound waves) associated to the spontaneous breakdown of local
space-time translations. The properties of the Goldstone modes are dictated by symmetry, and are therefore largely independent of the 
specific fields that gave rise to the background.

The paper is organized as follows.
In section~\ref{setup} we  describe the basic setup. In 
section~\ref{stability}
we study  the general conditions for stability.
We concentrate  on some unusual features of rotationally non-invariant
backgrounds, which make it rather
difficult to establish a relation between NEC violation
and instabilities in general. In particular, we explain that
the positive definiteness of the Hamiltonian is not  necessary
for stability. Another peculiarity is related to the
fact that in an effective field theory there is no {\it 
a priori}
reason to exclude the presence of  superluminal excitations around a
background which spontaneously violates Lorentz invariance. 
It turns out that
in the presence of  superluminal modes stability is not a frame
independent notion. More precisely a perfectly stable system may appear
to have ghost or gradient instabilities in some Lorentzian reference 
frames.

In section~\ref{proof} we establish a relation between NEC violation and
instabilities. We start by analyzing two-dimensional systems.  We show that if
Goldstone modes have non-degenerate dispersion relations\footnote{Degenerate
dispersion relations, where some modes have either vanishing gradient term or
vanishing time derivative term (or both), always emerge in the consistent
theories of massive gravity
\cite{Arkani-Hamed:2002sp,Arkani-Hamed:2003uy,Rubakov:2004eb,Dubovsky:2004sg}.}
and if the energy-momentum tensor can be diagonalized by a Lorentz boost, then
NEC violation implies the presence of an instability. On the other hand, we
construct a simple 2-dimensional model which is stable (the Hamiltonian is
even positive definite), violates NEC but possesses a non-diagonalizable
energy momentum tensor.
%a
%stable two-dimensional system (with a positive definite Hamiltonian) that
%violates NEC. A peculiar property of this system is that its
%energy-momentum tensor cannot be diagonalized by the Lorentz boost. 
We then present generalizations of these results to higher-dimensional
space-times. We prove that a violation of NEC leads to instabilities in a large class of situations. These include systems that are either rotationally invariant or possess only subluminal excitations.
 Nevertheless we were able to construct
an example of a stable low energy effective field theory that violates
NEC. This system has a positive definite Hamiltonian. Unlike in the  2-dimensional example,
the energy-momentum tensor of this system can be diagonalized by a Lorentz boost.
A peculiar necessary feature of  stable NEC violating systems is the
presence of  superluminal modes in the spectrum of perturbations.

In section~\ref{fluids} we provide a more general perspective on our
results. We relate scalar field theories with coordinate dependent condensates to  relativistic solids and fluids, and elaborate on this
correspondence, rederiving in field theory language some of the known results of hydrodynamics (such as Kelvin theorem). Although most of the results
of this section are well known, we find it useful to present this discussion
as it provides an intuitive physical picture of the systems considered in the
previous sections and in works on massive gravity. A pragmatic reader can skip this section
and proceed directly to  section~\ref{symmetries}.
In the latter   we study
NEC violation in  rotationally invariant systems with degenerate
dispersion relations. We argue that, unlike in  massive gravity \cite{Dubovsky:2004sg}, these
degeneracies cannot be protected by  symmetries.

In section~\ref{towards} we speculate on the implication of our results
were future more precise cosmological obervations to imply
 $w<-1$. 
We present simple arguments explaining why it is
difficult to mimic such an expansion in scalar-tensor theories of gravity without
introducing a NEC violating energy-momentum tensor. This allows us to propose a relatively
simple candidate model which can do the job without obviously contradicting  the precise tests of general relativity.
We summarize our results in section~\ref{summary}.

It is worth noting that a recent paper \cite{Buniy:2005vh} claims to have a proof
that NEC violation implies instabilities in a broad class of systems including
those considered in our work.  However, we believe that proof is not correct as we have been able to provide explicit counterexamples 
that contradict it (see section \ref{proof}).

\section{Setup}
\label{setup}
Let us consider a system of $N$ scalar fields
$\phi^I$, $i=1,\dots,N$ interacting with  gravity
\be
\label{action}
S=S_{gr}+\int \! d^4x \, \sqrt{-g} \, F(\phi^I,B^{IJ},\dots)\;,
\ee
where $S_{gr}$ is the gravitational part of the action,
\be
\label{BIJ}
B^{IJ}=g^{\mu\nu} \, \d_\mu\phi^I\d_\nu\phi^J\;,
\ee
and the dots stand for higher derivative terms. We work in the context of effective field theory
and assume that the scalar action has a cut-off  $\Lambda$ smaller than the
Planck scale $M_{Pl}$. 
%We assume that all fields $\phi^I$ are
%light compared to the cutoff scale $\Lambda$.
 More specifically we assume the function $F$ in eq.~(\ref{action}) is of the form
\be
\label{parametric}
F(\phi^I,B^{IJ},\dots)= \Lambda^4
\tilde{F} \Big({\epsilon\phi^I\over\Lambda},{B^{IJ}\over\Lambda^4},
{\d^2\phi^I\d^2\phi^J\over\Lambda^6}, \dots \Big)\;, \ee 
where $\epsilon\ll 1$ is a
small parameter responsible for the ``lightness'' of the fields $\phi^I$,
while all the other dimensionless parameters in the function $\tilde F$ are of
order one. Indeed for the very existence of a non trivial low energy effective
field theory the mass of the scalars $\sim \epsilon \Lambda$ should be much
less than the cut-off $\Lambda$.  Notice also that in the limit $\epsilon \to
0$ we would be dealing with a set of naturally massless Goldstone bosons.  We
want now to consider some classical solution $g_{cl}^{\mu\nu},\phi_{cl}^I$ and
study perturbations around it.  The validity of our effective field theory, or
the control over quantum corrections, forces us to consider only solutions
that are smooth on space-time scales of order $\Lambda^{-1}$
\be
|\d_\nu \phi^I |\gg \frac{|\d_\mu\d_\nu\phi^I |}{\Lambda},\dots \;.
\ee
In other words we assume that the fields do not vary appreciably over lengths
of order $1/\Lambda$ or more formally $|\partial _\mu | \ll \Lambda$ .  Notice
that this constraint does not necessarily imply\footnote{
In the Goldstone limit, $\epsilon=0$, the terms of the type $\left
[{B^{IJ}}\right ]^n$ represent the leading interaction among $n$-quanta in the
derivative expansion. In this respect $|B^{IJ}/\Lambda^4|$ is controlling the
size of classical non-linearities rather than the validity of the quantum
expansion, which is instead more directly tight to the condition $|\partial _\mu | \ll \Lambda$ 
(see a related discussion in Ref.~\cite{Nicolis:2004qq}).}
$\d_\mu\phi^I\ll\Lambda^2$.  Let us consider the relevant length scales. Since
the energy momentum generated by the scalars is of order $\Lambda^4$,
space-time will be curved over a length $r_G\sim M_{Pl}/\Lambda^2$. Under the
assumptions $\Lambda\ll M_{Pl}$ and $\epsilon \ll 1$ there is a range of lengths
\be \frac{1}{\Lambda} \ll r \ll r_G, \,\frac{1}{\epsilon \Lambda} \ee where we
can use effective theory, spacetime is approximately flat and the scalars can
be treated as massless.  In physically interesting situations where the
scalars describe some sort of cosmic fluid or a generalized quintessence, the
above range would encompass all relevant sub-horizon scales.  We want to study
perturbations around the solution in the above range of scales (cf., e.g.,
Ref.~\cite{Dubovsky:2005dw}), that is for space momenta $k$
%Let us study perturbations about this
%solution. We will focus on the behavior of these perturbations in a deep UV
%region  Namely, we consider
%a space-time region $\Gamma$ of a size $L$ much smaller than the typical
%masses of the fields and the curvature radius of the background.
%\[
%\Lambda^{-1}\ll L\ll (g\Lambda)^{-1}\;, {M_{Pl}/\Lambda^2}\;.
%\]
%The corresponding values of the momenta $p$ are
\be
\label{krange}
\frac{\Lambda^2}{ M_{Pl}},\;\epsilon\Lambda\ll k\ll\Lambda\;.  
\ee 
As we focus on a patch of space-time where curvature and scalar masses are negligible, by
 a suitable change of coordinates we can
approximate  $g_{cl}^{\mu\nu}$ by the Minkowski metric
$\eta^{\mu\nu}$ in the whole patch and the background scalar fields $\phi_{cl}^I$ by linear functions of
the coordinates
\be
\label{phisol}
\phi_{cl}^I=\phi_0^I+A_\mu^Ix^\mu\;.
\ee
Since we can neglect both masses and higher derivative terms we can define our function $F$ to be just
 $F\equiv \Lambda^4 \tilde{F}(B^{IJ}/\Lambda^4)$.
%For perturbations with the momenta in the range (\ref{krange}) one may neglect
%both non-derivative and higher-derivative terms in the action (\ref{action}).
%It is the existence of this window of momenta between the mass of the heaviest
%field and the cutoff scale which is important for our argument rather than a
%particular parametric form of the function $F$ in eq.~(\ref{parametric}).
Finally, at momenta larger than the curvature, $k\gg \Lambda^2/ M_{Pl}$, we can neglect the mixing between scalar fields and metric perturbations.   Indeed this mixing is a manifestation
of the gravitational Higgs mechanism: the mixing scalar modes
 are the Goldstone bosons of spontaneously broken translations, and the graviton has effectively a mass of the order of the curvature $\Lambda^2/M_{Pl}$. By a suitable
(unitary) choice of coordinates the Goldstones can be eliminated, in which case  extra physical polarizations appear in the graviton field. However when
working at $k\gg \Lambda^2/M_{Pl}$, the graviton mass can be neglected and, by an analogue of the equivalence theorem, the extra graviton polarizations can be accurately described  by the Goldstone bosons.
In the rest of the paper we will work in $\Lambda=1$ units.

By defining the perturbed scalar field as $\phi^I=\phi^I_{cl}+\pi^I$ we can write the  quadratic Lagrangian for  scalar perturbations in Fourier space as \be
\label{pertLagr}
L=L_{IJ}(p) \pi^I(-p)\pi^J(p)\;.  
\ee 
Here $p$ is the four-momentum and the matrix $L_{IJ}(p)$ is given by
\begin{gather}
\label{LIJ}
L_{IJ}=2F_{IK,JL}\, A^K_\mu p^\mu \, A^L_\nu p^\nu+F_{IJ} \, p^2\;,
\end{gather}
where
\begin{gather}
F_{IJ}={\d F\over\d B^{IJ}}\;,\;\;F_{IJ, \, KL}={\d^2 F\over\d B^{IJ}\d B^{KL}} \; .
\end{gather}
It is straightforward to check that the  background 
energy-momentum tensor $T_{\mu\nu}$ is given by
\be
\label{EMT}
T_{\mu\nu}=2F_{IJ}A_\mu^IA_\nu^J-Fg_{\mu\nu}\;.
\ee
Our goal is now to study the stability of eq.~(\ref{pertLagr}) when the background violates NEC, that is when there exists some light-like vector $n^\mu$ such that
\be
\label{NEC}
T_{\mu\nu}n^\mu n^\nu<0\;.
\ee

\section{General criteria for  stability}
\label{stability}
\subsection{One flavor in two dimensions}
\label{oneflavor}
Before addressing the stability of Lagrangian (\ref{pertLagr}) in 
full generality, it is instructive to discuss a simple special case: a single scalar field
 in (1+1)-dimensional space-time with action ($\dot \pi\equiv \partial_t \pi$, $\pi'\equiv \partial_x \pi$)
\be
\label{2dact}
S=\int \! dtdx \, \frac{1}{2}\l K \dot\pi^2+M \dot\pi\pi'-G\pi'^2\r\;.
\ee
The dispersion relation is
\be
\label{disp}
\lambda(\nu)\equiv K\nu^2+M\sigma\nu-G=0\;, 
\ee 
where
\be
\label{nu}
\nu=\omega/|k|
\ee
 is the ratio of  the frequency and the   absolute value of  the spatial
momentum $k$,
and $\sigma=\pm 1$ depending on the direction of
the spatial momentum. 

In general, the system (\ref{2dact}) can exhibit two different types of 
instability. First, there can be a gradient instability. This
is present whenever equation (\ref{disp}) has complex roots.  This
 is similar to the usual tachyonic instability which arises in a Lorentz invariant theory when some   field  has a negative mass squared, $m^2<0$. It is however different  since in the latter case the
time-scale $\tau$ with which the instability grows is set by the inverse
mass, $\tau\sim |m|^{-1}$. This time scale may be very long compared to the 
other characteristic time scales if $m^2$ is small enough. Equivalently, 
conventional tachyonic instabilities are present only for 
long wavelength modes.  On
the other hand if eq.~(\ref{disp}) has a complex root this implies the
presence of instabilities at all wavelengths, so that their rate of
growth is not bounded. Instabilities of this type are
significantly more dangerous than the usual tachyonic ones. Nonetheless in the rest of the
paper we will improperly refer to gradient instabilities as {\it tachyons}.

Another type of instability is the one associated with {\it ghosts}. In a
Lorentz invariant theory a ghost is a field entering the action with a wrong
sign kinetic term. At the classical level such a field has a negative definite
free Hamiltonian. Therefore, neglecting non linear terms or interactions with
other fields, no runaway behaviour occurs: there just happens to exist one
decoupled sector using the opposite sign convention for time and energy.
Runaway behaviour shows up when the ghost is set to interact with ordinary
fields having positive definite free Hamiltonian. Notice that, being due to
non-linearity, the time scale for growth becomes infinite in the limit of
vanishing initial perturbation, {\it i.e.}~in the classical vacuum. The situation is
generalized, but gets worst, at the quantum level, just because the field
fluctuates away from the origin already in the vacuum. The ghost quanta have
negative energy\footnote{We quantize on a physical Hilbert space with positive norm, so that
   all probabilities are non-negative.}, so that, once interactions are taken
   into account, the vacuum itself becomes unstable with respect to
   spontaneous creation of positive and negative energy quanta with vanishing
   total energy. Again, ghost instabilities are entirely due to the
   non-trivial couplings between negative and positive energy fields, and
   strictly speaking require to go beyond the quadratic approximation. We also
   recall that in a Lorentz invariant theory, the rate of development of such
   an instability is uncalculable (formally infinite), even for arbitrary
   small couplings, due to the non-compact final state phase space volume. A
   way to solve this problem is to introduce a cutoff which explicitly breaks
   Lorentz invariance. In such a way the rate of growth of the instability can
   be at least estimated and phenomenological consistency satisfied if the
   coupling between the ghosts and the other fields is small enough (see for
   instance refs. \cite{Cline:2003gs,Kaplan:2005rr}). In what follows we will
   not be concerned with the latter possibility but instead require the
   stronger, more robust, condition that ghost instabilities be absent.
%Since ghost instabilities are especially dangerous at the quantum 
%level, we refer to these instabilities as the quantum
%ones, but it is useful to keep in mind that the presence of ghosts leads
%to the classical instabilities as well (at the non-linear level the 
%system is 
%unstable towards exciting 
%modes of the positive and negative 
%energy fields with arbitrarily high energies).
As made clear from the discussion, whether or not negative energy states in
the system (\ref{2dact}) imply instability depends on the spectrum of the
fields coupled to them.  One coupling that surely exists is the one to the
graviton.  Moreover in the range we are focusing on, the graviton satisfies
to a good approximation a Lorentz invariant dispersion relation
$\omega^2=k^2$. We will therefore require the system to be kinematically
stable with respect to the associated production of $\pi$ quanta and of some
other massless particle (in particular the graviton)\footnote{Note, that in
two dimensions the graviton does not have propagating degrees of freedom. This
is not essential for our purposes as we will eventually be interested in four
dimensional theories.  In the 2d example we are therefore considering the
coupling to some other massless field.}.

Let us now study what the stability conditions are. Absence of tachyons requires 
 that both roots of the equation (\ref{disp}) be real, {\it i.e.}~ 
\be
\label{D>0}
M^2+4KG\geq 0\;.
\ee
The issue of  ghosts can be equivalently studied at the classical or quantum level.
We find it convenient to work right away at the quantum level and use the 
perhaps more intuitive particle physics viewpoint.  
The general solution of the field equations  following from the action (\ref{2dact})
has the  form
\be
\label{solution}
\pi(t,x)=\int \! dk  \l\alpha_{1}(k)\,  e^{ik(\nu_1 t-x)}+
\alpha_{2}(k) \, e^{ik(\nu_2 t-x)}\r\;,
\ee
where $\nu_{1,2}$ are the two solutions of  
eq.~(\ref{disp}) with $\sigma=1$.
Without loss of generality, we assume  $\nu_1>\nu_2$. 
Reality of the field $\pi$ implies
\be
\alpha_{1,2}(k)^\dagger=\alpha_{1,2}(-k)\;.
\ee
By eq.~(\ref{2dact}) the momentum $\Pi$ canonically conjugated to  $\pi$ is given by
\be
\Pi=K\dot\pi+\sfrac{1}{2}M\pi'\;.
\ee
The canonical commutation relations
%\[
%[\pi(x_1,t),\Pi(x_2,t)]=i\delta(x_1-x_2)
%\]
in momentum space become
\be
[\alpha_a(k),\alpha_b(k')]=\frac{2\delta_{ab}\sigma(K)(-1)^a}{\pi k\sqrt{M^2
+4KG}}\delta(k+k')\;,
\ee
where $\sigma(K)$ stands for the sign of $K$. 
It is convenient to consider separately the two cases $K>0$ and $K<0$.
\vskip1.0truecm

\noindent $\bullet \quad K>0 \,$. \\
In this case
we define creation and annihilation operators $a^\dagger(k)$, $a(k)$ as
\begin{gather}
a^\dagger(k)=\alpha_1(k)\theta(k)+\alpha_2(k)\theta(-k)\\
a(k)=\alpha_1(-k)\theta(k)+\alpha_2(-k)\theta(-k)
\end{gather}
Then, by defining the Fock vacuum as the state $|0\rangle$ satisfying
\be
a(k)|0\rangle= 0\; \quad\quad\forall k\,,
\ee
the Fock space generated by the action of the  $a^\dagger(k)$'s on $|0\rangle$ has a positive definite inner product. Consequently, the energy spectrum of
one particle states is
\be
E(k)=\nu_1k \, \theta(k)+\nu_2k \, \theta(-k)\;.
\label{energy}
\ee
Negative energy states  are absent provided $\nu_1$ is positive and $\nu_2$ is negative. For $K>0$ this condition is true if and only if $G>0$. Notice that positivity of $K$ and $G$ coincides with positivity of the classical Hamiltonian $H$ 
\be
\label{hamiltonian}
H= \int \! dx \, \frac{1}{2} \l K^{-1}(\Pi-\sfrac{1}{2}M\pi')^2+G\pi'^2 \r>0\;.
\ee
However, $K,G>0$ is not necessary for the stability of the
system (cf. Ref.~\cite{Nicolis:2004qq}). 
Indeed,  consider the situation where $\nu_1$ and $\nu_2$ are both positive, in which
eq.~(\ref{energy}) is not positive definite. 
The ghost instability is still absent, provided  the
two-momentum of the multiparticle states of the $\pi$-system (\ref{2dact})
never belongs to the interior of the lower light cone, that is if
\be
\label{lightconecond}
E>-|k|\;.  
\ee 
In this case, in spite of the presence of  negative
energy states, vacuum production of negative energy $\pi$-particles and
positive energy gravitons is not kinematically allowed. The basic reason for
stability in this case is that in addition to energy there exists an extra conserved charge, 
the total momentum.  Condition (\ref{lightconecond}) holds
even when both $\nu_1$ and $\nu_2$ are positive provided \be
\label{n2cond}
\nu_2<1\;,
\ee
or if  both $\nu_1$ and $\nu_2$ are negative provided
\be
\label{n1cond}
\nu_1>-1\;.
\ee 
In both cases the Hamiltonian is unbounded both from above and from below.
%Another way to see that there is no instability in these cases
%is to note that one can do a boost (in the direction of positive $k$ for
%positive $\nu_1,\nu_2$ and in the direction of negative $k$ for negative
%$\nu_1,\nu_2$), such that the energy spectrum is positive in a new reference
%frame, iff conditions (\ref{n2cond}), (\ref{n1cond}) hold.

\begin{figure}[t!]
\begin{center}
\includegraphics[width=13cm]{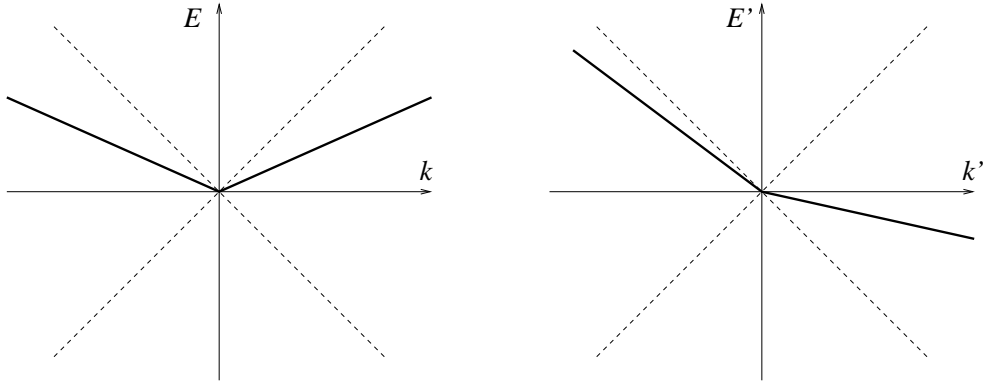}
\end{center}
\caption{\label{subluminalfig} The dispersion relations (bold lines) of a subluminal scalar in the
original frame in which eq.~(\ref{subluminal}) holds (left), and in a highly boosted frame (right).}
\end{figure}

To better understand the physics of such systems  let us consider a
subluminal scalar field in the reference frame where its action is
rotationally invariant\footnote{In two dimensions, rotations reduce to the 
 $\mathbb{Z}_2$ reflection $x\to -x$.},
\be
\label{subluminal}
S=\int \! dtdx \, \frac{1}{2}\l \dot\pi^2-v^2\pi'^2\r\;, \ee with $0<v^2<1$.  
The Hamiltonian associated to this action is positive definite and there is no
doubt that this system is perfectly stable.  Let us, however, consider
the same system as seen in  another Lorentzian reference frame, {\em i.e.}~let us perform a change of
time and space variables in the action (\ref{subluminal}) according to the
rule\footnote{Recall that
all our considerations are in the context specified in
section \ref{setup}: starting from a Lorentz invariant action, we study the dynamics of
fluctuations around a generic Lorentz-breaking field configuration.
Lorentz transformations are still a symmetry of the whole system.}
\be
\label{boost}
\l
\begin{array}{c}
t\\x
\end{array}
\r
\to
\l
\begin{array}{cc}
ch & sh\\
sh & ch
\end{array}
\r
\l
\begin{array}{c}
t\\x
\end{array}
\r\;,
\ee
where $ch=\cosh\lambda$ and $sh=\sinh\lambda$ are the boost parameters.
After this change of coordinates the action (\ref{subluminal}) takes the
general form (\ref{2dact}) with the coefficients $K$, $M$, and $G$ equal to
\begin{gather}
K=ch^2-sh^2v^2\nonumber\\
M=\sfrac{1}{2} \, ch \, sh \, (v^2-1)
\label{KMG}\\
G=ch^2 v^2-sh^2\nonumber
\end{gather}
One observes that for $v^2<1$ the coefficient $K$ remains positive in all
reference frames, while $G$ becomes negative for large enough boosts, 
$sh^2/ch^2>v^2$.  So, in the boosted frame, negative energy states appear and
the Hamiltonian is unbounded both from above and from below.  Dispersion relations
for one-particle states in the original and in the boosted reference
frames are shown in Fig.~\ref{subluminalfig}. It is clear from this figure
that  the conditions (\ref{n2cond}),
(\ref{n1cond}) hold in both frames as it is natural to expect for a stable
system.

Actually, the presence of  negative energy states is  easy to understand
in the following way.
Let us consider a massive particle with  relativistic dispersion relation
\be
E^2=k^2+m^2\;,
\ee
moving with  velocity $V$
 in the original reference frame. At $V>v$ the emission of  \^Cerenkov radiation
is kinematically allowed and 
the massive particle can emit one quantum of the $\pi$-field, thus losing  part of its
energy. On the other hand,
from the point of view of an inertial observer moving at velocity $V$,
 the massive particle is initially  at rest and then starts moving by emitting a  negative energy quantum.
\vskip1.0truecm

\noindent $\bullet \quad K<0 \,$. \\
Let us now come back to the study of the general action (\ref{2dact}) and
consider the case $K<0$.
It is straightforward to check that 
the energy spectrum of  one-particle states 
is now given by
\be
E(k)=\nu_1k \, \theta(-k)+\nu_2k \, \theta(k),
\ee
so that a state with two particles of equal and
opposite spatial momentum has  negative total energy (recall that $\nu_1>\nu_2$).
At first sight, this seems to necessarily imply that theories with
$K<0$ suffer from a fatal ghost instability.

\begin{figure}[t!]
\begin{center}
\includegraphics[width=13cm]{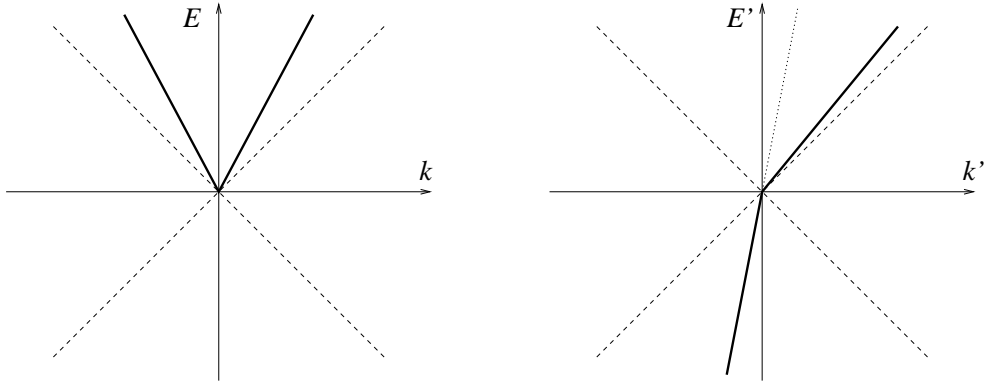}
\end{center}
\caption{\label{superluminalfig} The dispersion relations (bold lines) of a superluminal scalar in the
original frame (left) and in a highly boosted frame (right).}
\end{figure}

However, this naive conclusion is not quite correct. Indeed, let us again
consider action in  eq.~(\ref{subluminal}) but for the case of a
superluminal $\pi$ ($v^2>1$). There is no doubt that this action describes a
perfectly stable system in this case as well. Let us now look at this system
in the boosted frame (\ref{boost}).  eqs.~(\ref{KMG}) now imply that the
kinetic coefficient $K$ becomes negative for  large enough boost parameters,
${ch^2/ sh^2}>v^2$  (see Fig.~\ref{superluminalfig}).

\begin{figure}[b!]
\begin{center}
\includegraphics[width=13cm]{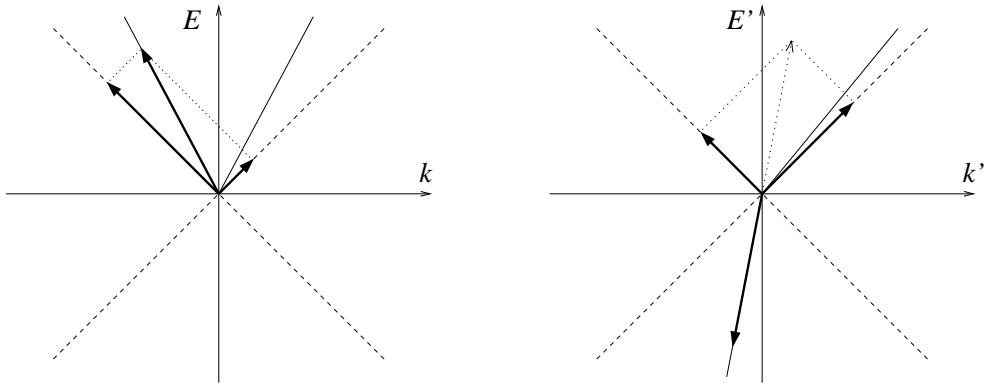}
\end{center}
\caption{\label{creation} The creation of a negative energy particle and two gravitons out of vacuum in the
boosted frame (right). The bold arrows are the particles momenta, which sum up to zero.
The same process appears in the original frame as the decay of a superluminal particle into two gravitons (left).}
\end{figure}

Naively, this result leads to a paradox where the vacuum of a
stable system (\ref{subluminal}) seems unstable to a fast moving observer.
Let us assume for the sake of the argument that  there exists
%for simplicity that there is no symmetry $\pi\to-\pi$ and 
a vertex with two gravitons and one $\pi$.
Then, compatibly with conservation of 2-momentum, the moving observer  expects to see events of spontaneous creation of one negative energy
$\pi$-particle and a pair of positive energy gravitons from the vacuum, see
Fig.~\ref{creation}.   He should then conclude that the vacuum is unstable.

\begin{figure}[t!]
\begin{center}
\includegraphics[width=13cm]{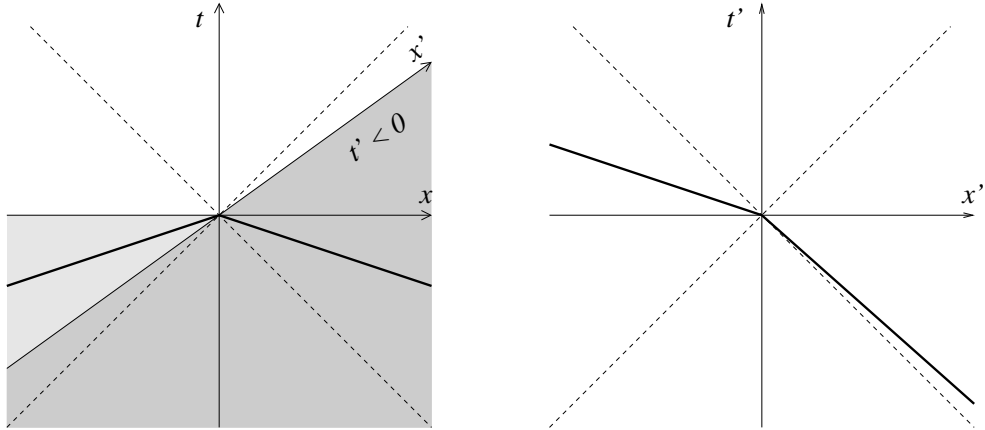}
\end{center}
\caption{\label{causalfig} The bold lines depict the past causal cones in the original reference frame (left) and in a highly boosted one (right).
Left: The two distinct vacuum choices require no incoming wave in the light shaded region ($t<0$) and in the dark
shaded region ($t'<0$), respectively. As clear from the picture these choices are inequivalent: an incoming particle moving along the left branch of the past causal cone is discarded in the former choice while it is accepted (as outgoing) in the latter.}
\end{figure}

The resolution of the paradox is just that the observer at rest and the moving
one have performed mutually inequivalent quantizations of the system.  This is
understood by considering the causal structure of the system and the initial
conditions associated to the two different parametrizations of time. First
notice that, for a large enough boost, part of the past causal cone is mapped
to the future of the new space-time coordinates. Consequently, an incoming
$\pi$-particle in the original (unboosted) reference frame, is mapped to an
outgoing $\pi$-ghost in the boosted frame. Therefore the decay process $\pi
\to 2g$ becomes the vacuum decay process $vacuum \to \pi+2g$ in the boosted
coordinates.  
Now the point is that the two observers have defined causality
by two {\it inequivalent} Cauchy surfaces, corresponding to surfaces of
constant time in the two different coordinate systems.  This is made evident
in Fig.~\ref{causalfig} where the two distinct vacuum choices correspond to
the request of no-incoming wave in two distinct shaded regions.  The two
choices are inequivalent because there are incoming (outgoing) portions of
particle trajectories that are allowed in one choice and not in the
other. Another way to say it is that the time vector, which defines the
hamiltonian flow (and its quantization), is inequivalently chosen in the two
parametrizations. This inequivalence is reflected by the fact that the
direction of time flow of one of the two particle trajectories that cross each
spacetime point is mutually reversed in the two cases. As shown in  Figure \ref{causalfig}
this is precisely like comparing two parametrizations of spacetime where the
role of time and space is interchanged\footnote{Notice, by the way, that if we
perform the exchange of time and space for an ordinary Lorentz invariant
massive field in two dimensions, we find that it becomes a tachyon.}.

In the end, as it should, the behaviour of the system does not depend on the observer but on the choice of quantization procedure, {\it i.e.}~on the vacuum state. Stability favors the obvious vacuum,
the one associated to the time slicing operated by the observer at rest.
Provided this is the vacuum, the moving observer will perhaps find it a little puzzling that he does not observe copious production of $\pi+2g$ from nothing.
However, with some thinking, he will realize that because of his unhappy choice of time coordinate, the correct causal past overlaps with a space time region with infinitely growing time coordinate.

 It is worth noting that there were
suggestions~\cite{Arkani-Hamed:2002fu} that
boundary conditions at  future
infinity may help to solve the cosmological constant problem.  It would be
interesting to understand whether the situation just discussed  may be relevant towards 
a realization  of such a proposal,  without invoking the anthropic principle and
without acausalities.

To finish this discussion, we remark that in the presence of superluminal
particles one can find examples where a stable system appears to have
a tachyonic instability in some Lorentzian reference frame.  It is
straightforward to check that this happens, for instance, if one adds a mass
term $-\frac12(m^2\pi^2)$ to the action (\ref{subluminal}) (for $v^2>1$).

To summarize, since the goal of the present paper is to find a connection
between a violation of the NEC and the presence of local instabilities, we
will adopt a very weak criterion of stability.  We will assume that if the
system (\ref{2dact}) possesses superluminal excitations one should not
conclude that it is unstable after observing (apparent) instabilities in some
specific Lorentzian frame. Instead one should check whether it is possible to
find some frame where the system (\ref{2dact}) is manifestly stable.  If this
frame exists, it should be used to quantize the system thus defining the
notion of causality\footnote{In the above discussion (and in the following
sections) we have restricted to Lorentzian frames the search of coordinate
where canonical quantization leads to stability.  It is easy to see that this
restriction is unimportant. As we explained, the choice of canonical
quantization corresponds to the choice of time coordinate $t$. Assuming a good
choice of equal time slicing exists, the invariant distance in that frame will
in general be $ds^2=adt^2+c_idt dx^i+g_{ij}dx^i dx^j$.  It is clear that by a
reparametrization of the form $t= c t'$, $x^i=c^i_j {x'}^j +c^i t'$ the metric
can be made Lorentzian without changing the time slicing.  }.  We leave aside
the issue whether an effective theory with superluminal excitations can arise
from a consistent microscopic theory.

% Assume we have found a frame where canonical q
%One may also ask why do we restrict our search for  stable frames to
%the Lorentzian frames. To justify this, let us show that if system (\ref{2dact})
%(coupled to the graviton) is stable in some frame, then it is possible to find a
%Lorentzian frame where it is manifestly stable as well.  Indeed, let us imagine that we
%found a frame where our system is manifestly stable and both the graviton and
%the $\pi$-field have actions of the general form (\ref{2dact}). Let us consider
%``pseudoLorentz'' transformations preserving the interval $(V^2dt^2-dx^2)$ with $V$
%larger than the propagation velocities of the graviton and 
%$\pi$-quanta in  both directions.
%According to the above discussion these transformations do not change the
%stability of the system. On the other hand one can check that using such
%pseudoboosts one can eliminate mixed term $\dot{h}h'$ in the graviton
%action. Afterwards one can perform a rescaling of the spatial coordinate to
%bring the
%graviton dispersion relation to the relativistic form $\omega^2=k^2$.

To conclude let us stress that in the absence of  superluminal excitations
the above peculiarity does not arise and all Lorentzian frames are equally
good for the analysis of  stability.

\subsection{Arbitrary number of flavors}
\label{manyflavors}
Let us now
 discuss what  the stability criterion is for a multiflavor system
of  general form (\ref{pertLagr}) with $I,J=1,\dots,N$. 
We still focus, for the moment, on two-dimensional systems. We work in the limit discussed in section \ref{setup}
in which  the kinetic matrix $L_{IJ}(p)$  is a quadratic form of momentum,
\be
L_{IJ}(p)=L_{IJ,\, \mu\nu}\, p^\mu p^\nu \,.
\ee
% for the most of our conclusions
%but it holds for the systems we are eventually interested in, so for 
%simplicity we stick to this case.
Dispersion relations for on-shell states are determined by the condition
\be
\det L_{IJ}(p)=0\;.
\ee
This equation defines a set of  rays in the
$(\omega,k)$ plane.   As before, we can rewrite it in terms of the slope (velocity)
$\nu$ as defined by eq.~(\ref{nu}),
\be
\label{Lnusigma}
\det \tilde{L}_{IJ}(\nu,\sigma)=0 \; ,
\ee
where $\tilde{L}_{IJ}(\nu)\equiv L_{IJ}(p)/k^2$ and 
$\sigma=\pm 1$ characterizes the direction of the spatial momentum $k$. 
Let us consider  eq.~(\ref{Lnusigma}) 
at $k>0$ \footnote{Note that $\tilde{L}_{IJ}(\nu,1)=\tilde{L}_{IJ}(-\nu,-1)$, 
so that $\tilde{L}_{IJ}(\nu,1)$ alone
carries all relevant information about the spectrum.}.
This equation has $2N$ roots. In analogy with the one flavor case, upon quantization
half of these roots $\{\nu_{+I}\}$ will correspond to  creation operators for  particles with 
momentum $k$ and  energy
\be
\label{nu+}
\omega=\nu_{+I}|k|\;.
\ee
The second half of the roots $\{\nu_{-I}\}$ corresponds to annihilation
operators for particles with momentum $-k$ and energy
\be
\label{nu-}
\omega=-\nu_{-I}|k|\;.
\ee
One necessary condition for   stability is that all roots $\nu_{\pm I}$
be real. 

To understand the issue of ghosts let us understand how the two sets, $\{\nu_{+I}\}$ and
$\{\nu_{-I}\}$, are characterized.  For this purpose, notice that (cfr.  the analysis in Ref.~\cite{Dubovsky:2004sg})  at any value of $\nu$, the real and 
symmetric matrix $\tilde{L}_{IJ}(\nu)$ has $N$ real eigenvalues 
$\lambda_I(\nu)$. When $\nu$ takes one of the values in $\{\nu_{+I}\}$ or in
$\{\nu_{-I}\}$ one of the eigenvalues $\lambda_I$ is equal to zero.
We will now prove that the two sets of zeros  are 
characterized by the sign of the derivative of the corresponding eigenvalue at
its zero,
\be
\label{+cond}
\left.\frac{d\lambda_I}{ d\nu}\right|_{\nu_{+J}}>0\mbox{ if }\;
\lambda_I(\nu_{+J})=0\;,
\ee
and
\be
\label{-cond}
\left.\frac{d\lambda_I}{ d\nu}\right|_{\nu_{-J}}<0\mbox{ if }\;
\lambda_I(\nu_{-J})=0\;.
\ee
%Before coming to the prove of this statement, let us note that 
%eqs.~(\ref{nu+})-(\ref{-cond}) imply that a solution $\nu_*$ of the dispersion
%relation corresponds to the positive energy state iff
%\[
%\left.\nu{d\lambda_I\over d\nu}\right|_{\nu=\nu_*}>0\mbox{ if }\;
%\lambda_I(\nu_{*})=0\;.
%\]
It is straightforward to check that these conditions agree with the results
of subsection~\ref{oneflavor} in the case of one flavor. 
%In principle
%one can prove conditions (\ref{+cond}), (\ref{-cond})  using the same method,
% {\it i.e.}~performing 
Conditions (\ref{+cond}), (\ref{-cond})  are easily derived by using the
K\"allen--Lehmann representation of the propagator.
Consider indeed the Feynman correlator
\[
G^{IJ}(t)=
\langle 0| \, T\left (\pi^I(t)\pi^J(0)\right )|0\rangle
\] 
and rewrite it by inserting a complete set of states $|n\rangle$ 
between $\pi^I(t)$ and $\pi^J(0)$
\be
\label{nn}
G^{IJ}(t)=\sum_n\theta(t) e^{-iE_nt}M^{IJ}_{(n)}+\sum_n\theta(-t) e^{iE_nt}M^{JI}_{(n)}\;,
\ee
where $E_n$ is the energy of the $n$-th state and
\be
\label{MIJ}
M^{IJ}_{(n)}=\langle 0|\pi^I(0)|n\rangle \langle n|\pi^J(0)|0\rangle\;.
\ee
Note that all the matrices $M^{IJ}_{(n)}$ are positive semi-definite. 
eq.~(\ref{nn}) implies that the Fourier transform $G^{IJ}(E)$ of the propagator
takes the  form
\be
\label{GE}
G^{IJ}(E)=\sum_n{i\over E-E_n+i\epsilon}M^{IJ}_{(n)}-\sum_n{i\over E+E_n-i\epsilon}
M^{JI}_{(n)}\;.
\ee
Consequently, the poles of $-iG^{IJ}(E)$ with positive residue correspond to the energies of the physical states.
On the other hand, from the Lagrangian we have
%at energy $E=E_n$  
%corresponding to the energy of the one-particle state $|n\rangle$
%is equal to $i M^{IJ}_{(n)}$. Furthermore,  the matrix $G^{IJ}(E)$ is related
%to the matrix $L_{IJ}(E)$ in the Lagrangian (\ref{pertLagr}) in a usual way,
\be
-iG^{IJ}(E)= L^{-1}_{IJ}(E) \;.
\ee
Consequently, using matrix notations we may write
\be
\label{OtO}
-iG(E)=k^2 O^T(\nu)D(\nu)O(\nu)\;,
\ee
where $D(\nu)$ is a diagonal matrix with  eigenvalues equal to $\lambda_I^{-1}(\nu)$
and $O(\nu)$ is an ortogonal matrix. Any pole of 
$G(E)$ corresponds to a zero of some 
$\lambda^I(\nu)$, while the sign of the residue is the same as the sign of the first derivative $d\lambda^I(\nu)/d\nu$. This
implies eqs.~(\ref{+cond}, \ref{-cond}).
%are the same as the sign
% and positive definiteness of the 
%matrix $M^{IJ}$ in eq.~(\ref{nn})
%implies 

Using these conditions one can determine the spectrum of the system.
According to our previous discussion, the system is stable provided we can so define time
that upon quantization there are no multiparticle states in the lower light cone (see 
eq.~(\ref{lightconecond})).
%In such a way,  spontaneous creation of $\pi$-quanta in association with ordinary massless particles,
%like the graviton, is not kinematically allowed.
\subsection{Arbitrary number of dimensions}
\label{manydim}
Let us finally discuss how the above results generalize to 
field theories in arbitrary  $(d+1)$ dimensions. A
straightforward generalization of the dispersion relation~(\ref{Lnusigma}) is
\be
\label{Lnuni}
\det \tilde{L}_{IJ}(\nu,{\bf \hat n})=0 \; ,
\ee
where, as before, $\tilde{L}_{IJ}(\nu,{\bf \hat n})\equiv L_{IJ}(p)/{\bf k}^2$ and 
${\bf \hat n}$ is now a unit $d$-dimensional vector in the  direction of the spatial momentum ${\bf k}$. 

Again the  system (\ref{pertLagr}) is stable if there is a Lorentzian 
reference frame, where
eq.~(\ref{Lnuni}) has 2$N$ real roots for any ${\bf \hat n}$ and there are no states
 in the
lower light cone. Note that our derivation of the conditions (\ref{+cond}),
(\ref{-cond}) was not sensitive to the number of the dimensions. Consequently,
 these conditions are applicable for arbitrary $d$ as well. 
The only difference is that now roots $\nu_{I \pm}$ are functions of the
unit vector ${\bf \hat n}$, and $\nu_{I +}({\bf \hat n})$-roots correspond to the particles with
momentum along ${\bf \hat n}$  while $\nu_{I -}({\bf \hat n})$-roots correspond to 
particles with momentum in the opposite direction.

\section{NEC violation and instabilities}
\label{proof}
To understand  the implications of the NEC violation for stability, 
let us first study how stability itself is encoded in the Lagrangian.
We start with the analysis of  two-dimensional systems. 
Consider a stable system of the type (\ref{action}).
According to the discussion of  section \ref{stability}
this means that a Lorentzian reference frame exists where the spectrum 
of states does not contain either tachyons or 
states with  energy-momentum in the lower light cone.
Let us see in more detail how this restricts the properties of the matrix
$L_{IJ}(p)$ in eq.~(\ref{pertLagr}).
As we discussed in subsection~\ref{manyflavors} the properties of the
spectrum can conveniently be read off from the properties of the eigenvalues 
$\lambda_I(\nu)$ of the matrix
\be
\tilde{L}_{IJ}(\nu)=L_{IJ}(p)/{k}^2\;,
\ee
where without loss of generality we assumed that $k>0$. Absence of tachyons implies that
these eigenvalues  have $2N$ zeros in total. Let us assume that we are in a 
generic situation, where all these roots are simple. 
Multiple roots may emerge due to some coincidental degeneracy or as the result of a symmetry. 
We postpone the 
discussion of   degeneracies to section~\ref{symmetries}.

\begin{figure}[b!]
\begin{center}
\includegraphics[width=15cm]{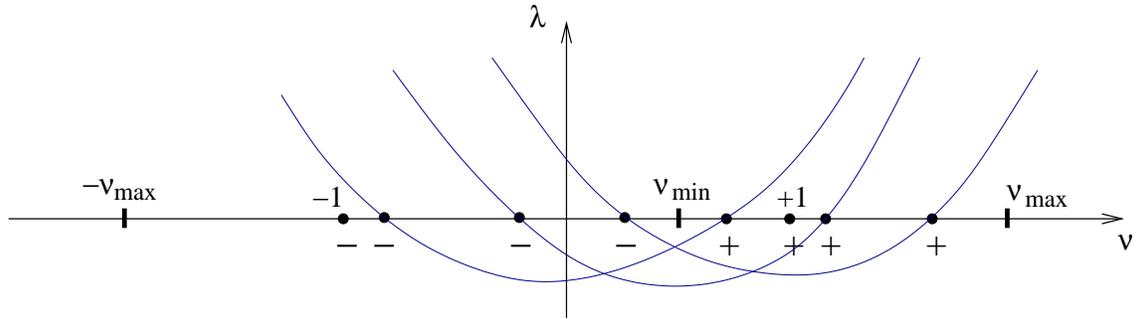}
\caption{\label{pm} The eigenvalues $\lambda_I$ of $\tilde{L}_{IJ}$ as functions of $\nu$.
The configuration shown corresponds to a stable system: all the positive derivative roots $\nu_{I+}$ are
on the right of the negative derivative roots $\nu_{I-}$ (taking into account also the roots $\pm1$ corresponding
to the dispersion relation of the graviton). As clear from the picture, this property is equivalent
to conditions {\it (i)}, {\it (ii)}.}
\end{center}
\end{figure}

Let us now understand what root configurations on the $\nu$-axis are implied
by stability. As before, we indicate by $\{\nu_{+}\}$ and $\{\nu_{-}\}$ the
sets of roots with respectively positive and negative slope.  In what follows
we consider all the particles together, so that we include in the sets
$\{\nu_{\pm}\}$ the roots $\nu=\pm 1$ associated to the ``graviton'',
``photon'', etc.  By eqs.~(\ref{nu+} -- \ref{-cond}), if there exist
$\nu_{1+}\in \{\nu_{I+}\}$ and $\nu_{1-}\in\{\nu_{I-}\}$ such that
$\nu_{1+}<\nu_{1-}$ then we can construct a two-particle state with zero total
momentum and negative energy
\be
E=(\nu_{1+}-\nu_{1-})k\;.
\ee
By the discussion in the previous section we thus have a ghost instability.
%We have thus a ghost instability when coupling the system to any 
%particle with Lorentz invariant disperision relation, {\it i.e} with $\nu=%\pm 1$. 
On the other hand it is straightforward to check that if  
\be
\label{ordering}
\nu_{I+}>\nu_{J-}
\ee
for any two members of the two sets 
(see Fig.~\ref{pm}), then the Hilbert space does not contain any state with  zero energy-momentum other than 
the vacuum, so that there is no ghost instability. 
%Consequently, 
%the only allowed configuration of pluses and minuses  is the one
%shown in Fig.~\ref{pm}a). Clearly,
eq.~(\ref{ordering}) is just a generalization of the conditions (\ref{nu+}), (\ref{nu-}) obtained in the  one flavor case.

Notice that eq.~(\ref{ordering}) implies that no eigenvalue $\lambda_I(\nu)$ has more than two zeroes on the $\nu$ line. 
Since we assume that there are $2N$ zeroes in total (we barr degeneracies), then each eigenvalue must have precisely two 
zeroes. The behaviour of the $\lambda_I(\nu)$ that follows is shown in Fig.~\ref{pm}. Since the signs of the eigenvalues 
define the sign properties of the matrix $L_{IJ}(\nu)$, we find that stability holds if and only if the following two 
conditions hold:
\begin{itemize}
\item[{\it (i)}] There exists $\nu_{\rm max}>1$, such that
$\tilde{L}_{IJ}(\nu)$ is positive definite for    $|\nu|>\nu_{\rm max}$.
\item[{\it (ii)}] There exists $-1< \nu_{\rm min}<1 $, such that   $\tilde{L}_{IJ}(\nu_{\rm min})$ is negative definite.
\end{itemize}
Condition {\it (i)} is just the usual requirement that the kinetic part of the Lagrangian (\ref{pertLagr}) be 
positive definite. Moreover if condition {\it (ii)} is satisfied for the special value $\nu_{\rm min}=0$, then 
the purely gradient part of the Lagrangian (\ref{pertLagr}) is negative definite.
Conditions {\it (i)}, {\it (ii)} with $\nu_{\rm min}=0$ are equivalent to positivity of the Hamiltonian. 
In general, however,  validity of {\it(i)} and {\it(ii)} is a weaker request than positivity of the Hamiltonian.

At this point it is convenient to go back to the matrix $L_{IJ}(p)$ and reformulate conditions  {\it (i)} 
and {\it (ii)} in a manifestly Lorentz covariant way. 
Let us define the following scalar function of 2-vectors $p$ and $q$  
\be
\label{Lpq}
L(p,q)\equiv L_{IJ}(p) \, A_\mu^Iq^\mu \, A^J_\nu q^\nu\;.
%=
%2F_{IK,JL}A^I_\mu A^K_\nu A^J_\lambda A^L_\sigma q^\mu p^\nu q^\lambda
%p^\sigma
\ee
In practice we are evaluating the quadratic Lagrangian of a perturbation
%The vector $q^\mu$ parametrizes perturbations  $\pi^I$ that
%belong to the following subspace of the whole vector space
\be
\label{qApert}
\pi^I=q^\mu A_\mu^I\;.
\ee
Perturbations of this special form can be obtained by acting on the vacuum (\ref{phisol}) 
with a space-time translation $x^\mu \to x^\mu +q^\mu$. 
Perturbations  (\ref{qApert}) are therefore
just the Goldstones (sound waves) of  broken space-time translations.
That is why we do not lose any relevant information by focusing on these perturbations. 
By using the defining eqs.~(\ref{LIJ}, \ref{EMT}), we obtain a simple relation
\be
\label{chain}
 L(p,q)-L(q,p)=\sfrac{1}{2}\l p^2 \, T_{\mu\nu}q^\mu q^\nu-q^2 \,
T_{\mu\nu}p^\mu p^\nu\r\equiv f(q,p, T_{\mu\nu})\;,
\ee
which relates the kinetic matrix to the energy momentum tensor, and which will prove
crucial in relating NEC to stability. As we shall see shortly, stability
in the form of conditions {\it (i)}, {\it (ii)} forces the function
$f$ to be positive for properly chosen $p^\mu$ and $q^\mu$. On the other hand, if 
NEC is violated then $T_{\mu\nu}$ is negative definite along some
direction. Our strategy will be to prove that under generic conditions there exists a tension between these two facts.

Now, notice that $|\nu|>1$ and $|\nu|<1$ correspond respectively to
time-like and space-like $2$-momentum.
Conditions {\it (i)}, {\it (ii)}, then  imply the following two properties of $L(p,q)$.
First, there exists a cone $\Gamma_+$ in the region of time-like momenta which includes the $(\omega \not = 0, k=0)$-axis
such that for any  $t^\mu\in \Gamma_+$ 
%which is sufficiently close to the $\omega$-axis and for an  arbitrary 
and for arbitrary  $q^\mu$
\be
\label{tmu}
L(t,q)>0\;.
\ee
Second, there exists a space-like line $\Gamma_-$ such that for $s^\mu\in \Gamma_- $ 
\be
\label{smu}
L(s,q)<0
\ee 
for any $q^\mu$.
%Let us see that conditions (\ref{tmu}), (\ref{smu}) cannot hold if the
%energy-mometum tensor is diagonalizable and violates the NEC.
By eqs.~(\ref{tmu}, \ref{smu}) it then follows that
\be
\label{chain2}
 f(s,t, T_{\mu\nu})>0\;,
\ee
for $t\in \Gamma_+$ and $s\in \Gamma_-$.

It is now straightforward to extend the discussion to systems in $d>2$.
Consider indeed the subspace of the states where all particles move collinearly ({\it i.e} the spatial momentum 
is along some  given direction ${\bf \hat n}$). Then one can literally repeat the arguments above and obtain that the 
conditions {\it (i)}, {\it (ii)} should hold for any direction ${\bf \hat n}$.
(In analogy with subsection \ref{manydim} the matrix $\tilde{L}_{IJ}(\nu)$ now 
depends on ${\bf \hat n}$ as well).
We stress however that these conditions, while necessary, are no longer  sufficient  to ensure the absence of ghost 
instabilities. 
Indeed, for anisotropic systems one may imagine a situation where states with  total momentum in the lower
light cone are absent if one considers only states with collinear particles, but are present in general.
Nonetheless, these necessary conditions allow us to establish a relation between NEC and stability in a rather 
broad class of theories. On the other hand we will be able to construct examples of stable systems 
that violate NEC. Therefore we do not see much motivation in trying to formulate stronger conditions.

\begin{figure}[t!]
\begin{center}
\includegraphics[width=14cm]{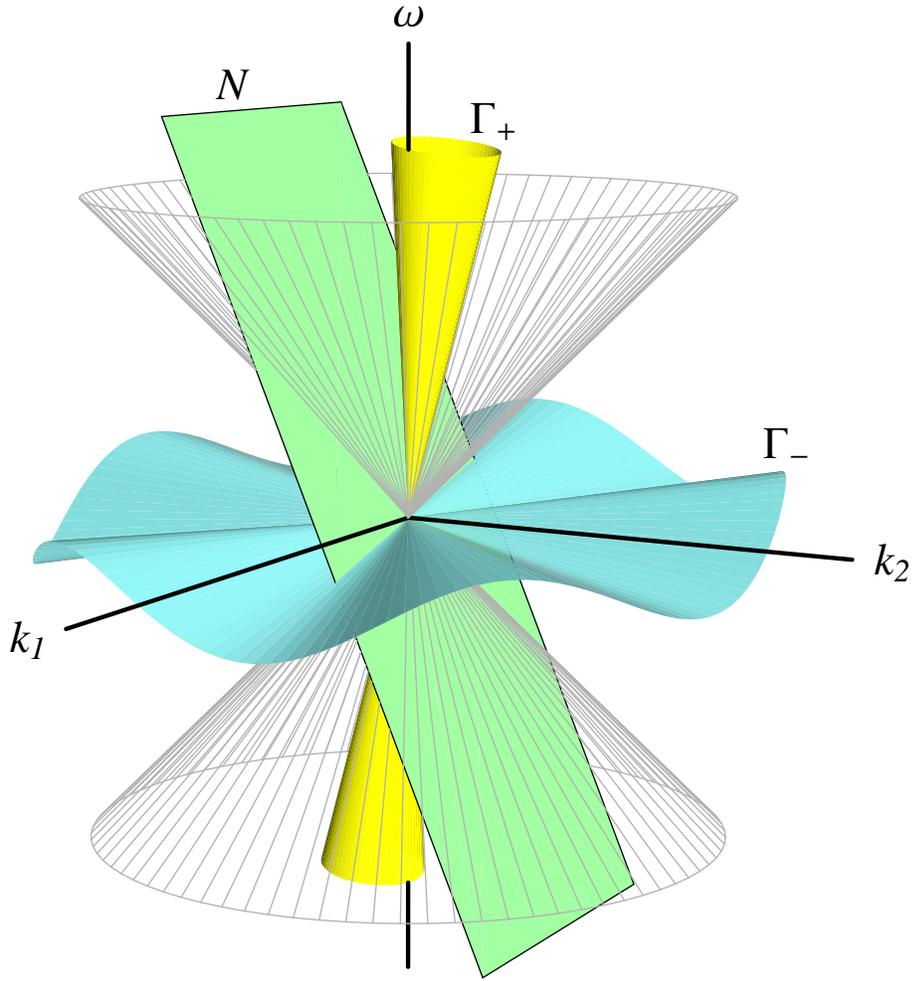}
\end{center}
\caption{\label{difficult_to_draw?} 
The various regions in the $(\omega,{\bf k})$ space discussed
in the text. $L_{IJ}$ is positive definite inside the yellow narrow cone
$\Gamma_+$, while it is negative definite on the skirt-shaped blue cone
$\Gamma_-$. The green plane ${\cal N}$ is one of the planes on which the
stress-energy tensor can be made negative definite. The grid-like cones
are the light-cones.}
\end{figure}

Focusing just on states of collinearly moving particles the conditions {\it (i)}, {\it (ii)} are easily generalized.
As a consequence condition (\ref{tmu}) is unchanged: there must exist a cone $\Gamma_+$  in  momentum space, 
which contains the $\omega$-axis and such that the matrix $L_{IJ}(p)$ is positive definite for any $p\in\Gamma_+$.
Condition (\ref{smu}) is similarly generalized: as before, for any given space direction ${\bf \hat n}$ 
there exists a line of space-like momenta along which the matrix $L_{IJ}(p)$ is negative definite.
By varying ${\bf \hat n}$ we find a skirt-shaped  cone $\Gamma_-$ (see fig.~\ref{difficult_to_draw?}).
As before stability implies that eq.~(\ref{chain2}) holds for $t\in \Gamma_+$ and $s\in \Gamma_-$.

Now that we have characterized the implications of stability, let us assume that the NEC condition
is violated along some null vector $n^\mu$: $T_{\mu\nu} \,n^\mu n^\nu < 0$. 
In $d>2$ by continuity there must exists a whole neighborhood of null directions along which NEC is violated. 
The two-dimensional case is singular in this respect, since there the light-cone is made up of two 
disconnected directions; we will analyze the implications of this fact later. Let us therefore concentrate on $d>2$ 
for the moment. 
Any pair of such linearly independent NEC-violating null vectors $n_1$ and $n_2$ defines a plane ${\cal N}$.
Now, both the NEC and the scalar quantity $f$ are invariant under the shift symmetry
\be
\label{shift}
T_{\mu\nu}\to \tilde{T}_{\mu\nu}=T_{\mu\nu}+\alpha \, \eta_{\mu\nu}\;,
\ee
for arbitrary $\alpha$. 
By choosing $\alpha = - T_{\mu\nu} \, n_1^\mu n_2^\nu / (n_1 \cdot n_2)$ we can make $T_{\mu\nu}$ strictly negative
definite on the whole plane ${\cal N}$, {\it i.e.}, on all linear combinations of $n_1$ and $n_2$.
Notice that ${\cal N}$ is a time-like plane: it intersects the light-cone along two lines (defined by $n_1$ and $n_2$ 
themselves), while a space-like plane would only intersect the light-cone at the origin.
By varying $n_1$ and $n_2$ inside the `NEC-violating neighborhood' we can construct a whole family of such time-like
planes ${\cal N}$ on which $T_{\mu\nu}$ can be made strictly negative by means of the suitable plane-dependent shift.
Then if we can prove that at least one of these ${\cal N}$'s must intersect both $\Gamma_+$ and $\Gamma_-$ 
we have a contradiction. Indeed,
 on the one hand stability forces $f(s,t,T_{\mu\nu})$ to be positive for all 
$s\in \Gamma_-$ and $t \in \Gamma_+$. On the other hand, if NEC is violated and one of the ${\cal N}$'s 
intersects both $\Gamma_+$ and $\Gamma_-$ we can pick $t^\mu$ and $s^\mu$ along these intersections. 
In this case $T_{\mu\nu}$ is negative definite both on $t^\mu$ and on $s^\mu$, and $f(s,t,T_{\mu\nu})$ is thus negative, 
by its very definition eq.~(\ref{chain2}).

It is evident that all $\cal N$'s must intersect $\Gamma_-$, since the former are time-like planes while the latter
is a space-like hypersurface. On the other hand they may not intersect $\Gamma_+$. 
This clearly happens when, in the frame where stability is manifest,
all ${\cal N}$'s are almost tangent to the light cone (which is the case if the neighborhood of NEC-violating null 
vectors is very small) and $\Gamma_+$ is a tiny cone around the energy axis. 
This suggests that in order to establish a connection between NEC and stability we need extra assumptions. 
These are indeed offered in at least two physically relevant situations:
\begin{itemize}
\item 
The first is the case where all excitations are subluminal, for which the region $\Gamma_+$ necessarily contains 
the light cone. Then all planes $\cal N$ intersects $\Gamma_+$ and we obtain that the assumption of stability and 
NEC violation contradict each other. 
\item
The other example consists of systems in which superluminal modes may be present but there are two 
NEC violating null vectors with opposite spatial components $n^\mu_{1,2}=(1,\pm {\bf \hat n})$. In this case
the plane ${\cal N}$ generated by $n_1$ and $n_2$ contains the $\omega$-axis and necessarily intersect the
$\Gamma_+$-region. This property trivially holds for systems, where the NEC is violated for all null vectors, 
in particular for isotropic systems. This is the case which is relevant for cosmological applications.
\end{itemize}

The fact that NEC follows from stability and subluminality, is perhaps not completely surprising. As discussed for instance if ref. \cite{Wald}
the dominant energy condition, which implies NEC, corresponds  to
a background with positive energy density and subluminal energy flow.
However the energy density and flow of the background do not precisely correspond to the Hamiltonian and velocity of the fluctuations around the background itself.  Mathematically this is seen by
comparing eq.~(\ref{LIJ}) and eq.~(\ref{EMT}). The matrix describing the propagation of perturbations involves also the second
derivative tensor $F_{IJKL}$ while the energy momentum does not.
Conversely $T_{\mu\nu}$ involves the cosmological constant term $\propto g_{\mu\nu}F$ which does not affect the propagation of the modes. These differences suggest that a quick conclusion that stability and subluminality of the modes implies NEC would be too naive. However,
 the quantities we have used in eq.~(\ref{chain}) somewhat
cleverly eliminate the contributions of $F_{IJKL}$ and $F$ in such
a way that the naive conclusion is correct.

Let us now come back to the two-dimensional case. In $d=2$ the light-cone is composed of the
two null directions $n^\mu_\pm=(1,\pm 1)$. If NEC is violated along both null directions, 
then, as before, the stress-energy tensor can be made negative definite on the plane ${\cal N}$ 
generated by $n_+$ and $n_-$, which in this case coincides with the whole $(\omega,k)$ plane!
This plane obviously intersects (in fact contains) both $\Gamma_+$ and $\Gamma_-$. 
We are therefore led to the above contradiction.
If instead NEC is only violated along, say, $n^\mu_+$, then the region ${\cal N}$ on which the 
stress-energy tensor can be made negative definite is just the line generated by $n^\mu_+$. 
In this case ${\cal N}$ cannot intersect both $\Gamma_+$ and $\Gamma_-$, and there is no evident contradiction.
A peculiar feature of two-dimensional systems is that the condition that NEC be violated along one and 
only one null direction is equivalent to the requirement that the energy-momentum tensor cannot be diagonalized by 
a Lorentz-boost. Indeed they are both equivalent to the inequality
\be
\label{nond}
|T_{00}+T_{11}|<2|T_{01}| \;.
\ee 
We can therefore summarize the $d=2$ case as follows. If the stress-energy tensor is diagonalizable by a 
Lorentz boost and the NEC condition is violated, then the system is unstable.
On the other hand we will now show that by giving up the assumption that $T_{\mu\nu}$ be diagonalizable
we can construct a system that violates NEC while being perfectly stable, indeed with a positive Hamiltonian.
We stress again that such an assumption is only crucial in $d=2$, given its accidental relation to the `extent'
at which the NEC is violated. Indeed in the next subsection we will also show an example in $d>2$ of a stable 
NEC violating system with diagonalizable energy-momentum tensor.

\subsection{Examples of stable, NEC-violating systems}
\label{examples}
In $d=2$, consider two scalar fields $\phi^I$, $I=1,2$ with a purely derivative action
and choose the following classical background
\be
\label{tx}
\phi^0=t , \qquad\phi^1=x \,.
\ee
By defining
\be
\label{tildeBIJ}
\tilde{B}^{IJ}\equiv B^{IJ}-\delta^{IJ}
\ee
(with the ``meson'' field  $B^{IJ}$ defined in eq.~(\ref{BIJ}))
we now present an action written in term of $\tilde{B}^{IJ}$
such that $T_{\mu\nu}$ satisfies eq.~(\ref{nond})  and such that the quadratic Hamiltonian of small perturbations  
is positive definite. The advantage of the definition
(\ref{tildeBIJ}) is that $\tilde{B}^{IJ}=0$ on the background (\ref{tx}).
This simplifies some of the algebra. The Lagrangian is of the form
\be
\label{2dexample}
{\cal L}= \left(\tilde B^{00}+ \tilde B^{11}+2(1+a^2)\, \tilde B^{01}-(a^2+b^2)\, \tilde B^{00} \tilde B^{01}+
(a^2+c^2)\, \tilde B^{11} \tilde B^{01}\right) \;.
\ee
with \[
b^2 <2,\qquad c^2<2\;.
\] 
 
The energy-momentum tensor for this system can be calculated using equation \eqref{EMT},
\be
T_{\mu \nu} = \begin{pmatrix} 1 && 1+a^2 \\ 1+a^2 && 1 \end{pmatrix}
\ee
which satisfies \eqref{nond} and violates the NEC. To study the stability of the system we need to write the Lagrangian for the fluctuations and this is done in the appendix.

To conclude this example, let us note that this construction can be 
readily extended to provide an example of a stable system with 
non-diagonalizable (and, consequently, violating NEC)
energy-momentum tensor
for an arbitrary number of dimensions $d$. In addition to  $\phi^0$,
$\phi^1$ one introduces more scalar fields 
$\phi^i$, $i=2,\dots,d-1$ with background values
\[
\phi^i_{cl}=x^i \;.
\]
As before we will denote small perturbations of the $\phi^i$-fields
about this background by $\pi^i$.
We also add the following term to the Lagrangian (\ref{2dexample}) 
\be
\label{2dmany}
\Delta{\cal L}=\left(N^2\tilde B^{ii}-f^2  \big[ (\tilde B^{0i})^2+(\tilde B^{1i})^2 \big]\right)
\ee
where we generalized the definition of the $\tilde B_{ij}$ in the obvious way.
For large enough $f^2$ the last two terms in eq.~(\ref{2dmany})
provide a dominant
contribution of the correct sign to  the gradient terms of the $\pi^{0,1}$-fields in the  $x^i$-direction. 
The first term in  eq.~(\ref{2dmany}) provides healthy kinetic and gradient terms for the $\pi^i$-perturbations. 
In the limit $N^2\gg a^2$, $b^2$, $c^2$, 1 one may neglect extra mixings  involving $\pi^i$-fields and coming from the last two terms.

As anticipated, we now want to show that the requirement
 that $T_{\mu\nu}$ cannot be diagonalized by a Lorentz boost
is not crucial at $d>2$. Indeed we construct an example of a stable NEC violating system with
diagonalizable energy-momentum tensor.
We work in $d=3$, the generalization to $d>3$ being straightforward.
As a first step, we take the stress tensor, in the frame where it is diagonal, to be of the form
\be
T^{{\rm (diag)}}_{\mu\nu}=
\l\begin{array}{ccc}
a^2 & 0& 0\\
0& -b^2& 0\\
0& 0& 1
\end{array} 
\r 
\ee  
with $(a^2-b^2)<0$. 
By the discussion in the previous section the reference frame where a system 
with such $T_{\mu\nu}$ has a chance of being manifestly stable, 
must be boosted in such a way that all the null NEC-violating vectors are close to each other and the ${\cal N}$-planes 
they define are thus almost tangent to the light cone. 
The above stress-energy tensor violates NEC for null vectors whose spatial component mainly extends in 
the $x^1$-direction.
We thus consider a large boost in the $x^2$-direction.
In the boosted frame the energy-momentum tensor becomes
\be
\label{boostedT}
T_{\mu\nu}=
\l\begin{array}{ccc}
ch^2a^2+sh^2 & 0& ch \, sh \, (1+a^2)\\
0& -b^2& 0\\
ch \, sh \, (1+a^2)& 0& ch^2+sh^2a^2
\end{array}
\r          
\ee  
where $sh=\sinh\lambda$ and $ch=\cosh\lambda\gg a^2,\; b^2,\;1$,  are the boost parameters.
To construct a stable system  with stress tensor given by eq.~(\ref{boostedT}), we introduce three scalar field $\phi^I$ with vev's 
\be
\label{vacuum}
\phi_{cl}^I=x^I \; .
\ee
and write the Lagrangian in term of the mesons $\tilde{B}^{IJ}$ defined previously
\begin{multline}
\label{finalaction}
{\cal L} = T_{IJ} \, \tilde B^{IJ}+ \left(\alpha^2 \big[(\tilde B^{01})^2-(\tilde B^{11})^2-(\tilde B^{12})^2 \big]
+ \right. \\
\left. ch\, sh \, (1+a^2) \big[\tilde B^{02}(-\tilde B^{11}+\tilde B^{22})+ 2\tilde B^{01}\tilde B^{12} \big] \right) \;
\end{multline}
 with  $a^2\sim b^2\sim\alpha^2\sim 1$.
The stability of this system is discussed in the appendix. 
In agreement with our general expectations it contains a highly superluminal mode.

%%%%%%%%%%%%%%%%%%%%%%%%%%%%%%%%%%%%%%%%%%%%%%%%%%%%%%%%%%%%%%%%%%%%%%%%%%%%%%
%%%%%%%%%%%%%%%%%%%%%%%%%%%%%%%%%%%%%%%%%%%%%%%%%%%%%%%%%%%%%%%%%%%%%%%%%%%%%%

\section{Effective field theory for solids and fluids}
\label{fluids}

In the previous sections we focused our attention on systems involving derivatively
coupled scalar fields. 
It is not obvious a priori whether our results can be exported to different physical systems.
In particular, can we apply them to fluids? After all, most of the cosmic history is dominated by a mixture of 
fluids, their physical relevance in cosmology certainly overwhelms that of scalar fields.
In this section we show that our results are straightforwardly applied to universe filling
continua, solids, and in particular to
perfect fluids. Indeed the dynamics of these systems is described by an effective 
theory of three derivatively coupled scalars, as we will now discuss. Although
 most of the results we will discuss are well known,
we believe our presentation to be still useful as it provides a direct connection to the subject of the previous
sections and to the investigations on modifications of gravity.
For a recent thorough review on the Lagrangian formulation of fluid dynamics we refer to Ref.~\cite{jackiw}. Our treatment
is closer to the effective (non-relativistic) theory for supersolids developed in Ref.~\cite{son},  but we work from the beginning with the fully relativistic formulation.

\subsection{Generalities}

Let us consider a mechanical continuum filling all of (3-dimensional) flat space and such that at
each space point there resides just one (and only one) dynamical entity. There thus  exists
an invertible mapping $\phi$ between coordinate space and the internal space $\cal M$
labeling the dynamical variables (for instance the positions of the atoms in a crystal). This
is made explicit by working with a field theory of three scalar fields $\phi^I(x,t)$, $I=1,2,3$, which describe both  the mapping and the dynamics. Since $\cal M$ is diffeomorphic to $\mathbb{R}^3$ we can always choose
field coordinates $\phi^I$ such that in some given stationary state, for instance the ground state itself, the mapping is just $\phi^I(x^i, t) = x^I$. In what follows we shall indeed assume  $\phi^I(x^i, t) = x^I$ on the ground state.  Notice also that the $\phi^I$'s correspond to a choice of space coordinates comoving with the solid. 
This is to say that, given an arbitrary time dependent configuration $\phi^I(x^i,t)$, we can choose space-time coordinates by
\be
{x'}^i = \phi^i(x,t)\quad\quad t'=t
\ee
for which the scalar fields take constant value 
\be
{\phi'}^I (x',t')= \phi^I(x(x'),t') = {x'}^I \,.
\ee
Of course in the primed coordinates all the information about the state of the solid is contained in the metric,
which is no longer Minkowskian.

Let us work in Minkowskian coordinates around the ground state $\phi^I(x^i, t) = x^I$, and let us consider the symmetries of the system.
Ordinary space translations $T_x:\, \phi^I(x,t) \to \phi^I(x+a,t)$ are obviously a (spontaneously) broken symmetry.
On the other hand we would like to focus on {\it homogeneous solids}, for which there still
exists an unbroken translational invariance.  It is evident that the only way to achieve that,
is to further assume the additional internal shift symmetry
\be
T_\phi: \quad\phi^I(x,t)\to\phi^I(x,t)+c^I\,.
\ee
This way the ground state is invariant under a combination of ordinary translations and $T_\phi$ with 
$a^I=-c^I$. Therefore $T_x-T_\phi$ represents the conserved three momentum for processes
corresponding to small fluctuations around the vacuum (for instance in the scattering of phonons).
The broken generators $T_x+T_\phi$ are instead in one to one correspondence, by Goldstone
theorem, with 3 gapless excitations, the phonons. At the lowest derivative level, the relativistic invariant
Lagrangian for such a system will clearly be written as a function of the `meson' fields $B^{IJ}=\partial_\mu\phi^I\partial^\mu \phi^J$. It thus falls precisely into the class of field theories we have been discussing
throughout the paper! If the Lagrangian is a generic function of the $B^{IJ}$, the system will
describe an homogeneous but otherwise anisotropic solid. More symmetric systems can be constructed
by assuming additional internal symmetries acting on $\phi^I$. For instance we can consider an invariance
\be
\phi^I \to D^I {}_J \phi^J
\ee
where $D^I {}_J$ belong to some discrete subgroup of $O(3)$. In this case our system will have the same
macroscopic properties as a crystal with the corresponding symmetry. We can also consider
the limiting case in which  $D^I {}_J$ is extended to the full rotation group $SO(3)$. In which case
our system will essentially describe a jelly. 
%Let us focus on the latter case. Because of rotational invariance,
%the derivatives defined in section \ref{setup} have the form
%\be
%F_{IJ}=\alpha \delta_{IJ} \quad\quad F_{IK,JL}=\beta( \delta_{IJ}\delta_{KL}+\delta_{IL}\delta_{KJ})+\gamma \delta_{IK}\delta_{JL}
%\ee
%while $A^I_i=\delta ^I_i$ and  $A^I_0=0$.
%For this system it is straightforward to check that the absence of ghosts implies the NEC. Indeed we have
%\be
%\rho \equiv T_{00}= -F\quad\quad p\delta_{ij}=T_{ij}= (F -2\alpha)\delta_{ij}
%\ee
%so that $\rho +p=-2\alpha$.
%On the other end the quadratic action for the perturbation has the form
%\be
%-2\alpha \dot \pi^I\dot\pi^J\delta_{IJ}+\{{\rm terms\, with\, space\, derivatives\, only }\}
%\ee
%for which absence of ghosts implies $\alpha<0$ so that $\rho+p>0$.

The jelly system
% for which the internal symmetry corresponds
%to the group of roto-translations $\phi^I\to O^I_J \phi^J +c^I$, with $O_J^I\in O(3)$
does not correspond yet to the most symmetric
possibility. Indeed one can consider a system for which this
symmetry is extended to the group of volume preserving diffeomorphisms
\be
\label{noshape}
\phi^I\to \hat\phi^{I}(\phi), \qquad {\rm det} \, \frac{\d \hat\phi^{I}}{\d
\phi^J} = 1 \; .  \ee While this is a huge internal symmetry it still allows a
non trivial invariant quantity: the determinant of $B^{IJ}$. The most general
Lagrangian for such a system is then \be \label{S_fluid} {\cal L} = F(B) \; ,
\qquad B \equiv {\rm det} \, B^{IJ} \; , \ee where $F$ is a generic
function. It is pretty obvious that, with three fields, we cannot further
extend the volume preserving diffeomorphism to a larger symmetry that still
allows for a non-trivial action. What is the physical interpretation of a
system of this type? The answer is intuitively obvious: our Lagrangian
describes the dynamics of a (relativistic) fluid. Consider indeed the ground
state of the system $\phi^I(x)=x^I$.  We can imagine, starting from this
configuration, to build another one by moving around the individual points:
$x\to f(x)$. In the new configuration the field profile will be given by \be
\label{volumediff}
\phi^I(x)= (f^{-1})^I(x)\,.
\ee
Now, the intuitive notion of fluid suggests that its energy does not change  if we do
not compress it, no matter what different form its individual parts assume. Therefore
we expect that for a coordinate transformation $f$ that preserves the volume in physical space, the energy of the system should remain the same. This is obviously true
in our case because eq.~(\ref{volumediff}) can be compensated by an internal symmetry 
transformation.
%\footnote{It is amusing that
%our everyday's life experience that water has no shape is formally reflected in a local symmetry
%in field space, eq.~(\ref{noshape}).
% See also {\it La forma dell'acqua} by A. Camilleri....}$\phi^I\to f^I(\phi)$.

\subsection{Fluids}
In this subsection we want to work out in some detail the properties of the relativistic fluid
that follow from the Lagrangian in eq.~(\ref{S_fluid}). We want first to make contact with the standard
description that can be found, for instance, in Ref.~\cite{weinberg}. In particular we want
to find the field theoretic description of quantities like the fluid 4-velocity, the current, the pressure and energy density.

The definition of the $\phi$'s as comoving coordinates makes the
characterization of the fluid velocity field $u^\mu (x)$ immediate: $u^\mu$
is a vector field along which all the $\phi^I$'s stay constant,
\be \label{constraint_u}
\frac {d }{d \tau}\phi^I (x(\tau)) \equiv u^\mu \, \d_\mu \phi^I = 0 \qquad
I=1,2,3 \; . 
\ee
These orthogonality constraints together with the normalization constraint $u^\mu
u_\mu = -1$ completely determine $u^\mu$ (except for the discrete
choice of the direction, which is set by the request that $u^\mu$ be
future-directed).
Indeed the orthogonality constraints forces $u^\mu$ to be proportional
to the `vector product' $\epsilon^{\mu\alpha\beta\gamma} \, \d_\alpha \phi^1
\d_\beta \phi^2 \d_\gamma \phi^3 \propto
\epsilon^{\mu\alpha\beta\gamma} \, \epsilon_{IJK} \, \d_\alpha \phi^I
\d_\beta \phi^J \d_\gamma \phi^K$.
The proper normalization is
\be
u^\mu = \frac 1 {6 \sqrt{B}} \,\epsilon^{\mu\alpha\beta\gamma} \, \epsilon_{IJK} \, \d_\alpha \phi^I
\d_\beta \phi^J \d_\gamma \phi^K \; .
\ee
Note that this expression is invariant under generic (not necessarily
volume-preserving) diffeomorphisms acting on the $\phi$'s. This is
correct, since what matters for characterizing  $u^\mu$ is only that
the $\phi$'s are comoving coordinates. By the above equation, $u^\mu$
is associated to the volume form $\Omega$ over the fluid space
\be
u_\mu dx^\mu = \frac{1}{\sqrt B} \star( \sfrac{1}{6} \epsilon_{IJK} d\phi^I\wedge d\phi^J\wedge d\phi^K)\equiv 
\frac{1}{\sqrt B} \star \Omega
\ee
and the one form $\star \Omega\equiv J_\mu dx^\mu $ is naturally interpreted as the current of fluid points. Consistent with this interpretation, the current is conserved
\be
d \star(J_\mu dx^\mu)= d (\sfrac{1}{6} \epsilon_{IJK}d\phi^I\wedge d\phi^J\wedge d\phi^K) = 0\, .
\ee
Notice that the conservation of $J_\mu$, rather  than arising on-shell from a symmetry of the Lagrangian via Noether's theorem, is already valid  off-shell since the number of degrees of freedom, that is the volume of $\cal M$, is fixed by construction. 
From $J_\mu = n \cdot u_\mu$ it also follows that the density of particles $n$ is
\be
\label{n}
n = \sqrt B
\ee

What about the equation of state? 
%Everything must be encoded in the action, {\it i.e.}~in $F(B)$. 
%So is the equation of state.
%In particular it can be read from the stress-energy tensor.
The stress-energy tensor is given by eq.~(\ref{EMT}), with $F_{IJ} =
\frac{\d F}{\d B^{IJ}} = F' \, B \, B^{-1} _{IJ} $,
\be \label{fluid_Tmn}
T_{\mu\nu} = 2 F' \,  B \, B^{-1} _{IJ} \: \d_\mu \phi^I \d_\nu \phi^J
- g_{\mu\nu} F \; .
\ee
On the other hand for a perfect fluid the general expression is
$T_{\mu\nu} = (\rho + p) \, u_\mu  u_\nu + p \, g_{\mu\nu}$.
The energy density and pressure therefore can be extracted by contracting
$T_{\mu\nu}$ with $g^{\mu\nu}$ and with $u^\mu u^\nu$. By making use
of the constraints eq.~(\ref{constraint_u}) we get $\rho$ and $p$ as
functions of $B$,
\be
\label{rhop}
\rho  = -F(B) \; , \qquad p  =-2 F'(B) \, B +F(B) \; .
\ee
Note that although the stress-energy tensor eq.~(\ref{fluid_Tmn}) for a generic configuration depends on the full structure 
of $\d_\mu \phi^I$, the energy density and pressure depend on $B$ only. This is physically correct: like the Lagrangian they must be scalar quantities only sensitive to compressional modes.

By eqs.~(\ref{n}, \ref{rhop}), the three quantities  $n,p,\rho$ are determined by just one parameter.
On the other hand, the equation of state of a fluid corresponds to just one constraint $f(n,p,\rho)=0$.
The additional local constraint arises, in our description, from the homogeneity of the initial entropy density together with the isoentropicity of the fluid transformations, as we now explain.
 In the ordinary treatment, see \cite{weinberg}, the current and energy momentum conservation equations
 ($\partial_\mu J^\mu=\partial_\mu T^{\mu\nu}=0$) can be combined to give an equation that establishes
 the conservation of the entropy per comoving volume $\sigma$ along the flux
\be
0=u^\mu\partial_\mu \sigma=\frac{1}{kT}\Big\{p \, \partial_\mu (1/n)+\partial_\mu(\rho/n)\Big\} u^\mu
\ee
In other words the motion of each fluid portion is isoentropic, although the entropy density $\sigma$ varies
in general through space. Therefore the above equation introduces an additional, non-algebraic---{\em i.e.},
differential---constraint
between $n,p,\rho$. Now, the fluid we are describing by our field theory is homogeneous from all stand points, in the sense that any given configuration can always be brought to the homogeneous ground state by means of {\em mechanical} transformations. In other words, in our formalism there is no room for genuine thermodynamic degrees of freedom: to increase the temperature we have to compress the fluid, we cannot `heat it up'.   
In particular our fluid is homoentropic (the entropy per comoving volume is the same for all fluid elements), in which case the last equation is replaced by the stronger one $\partial_\mu \sigma = 0$ which corresponds precisely to the additional algebraic constraint we have between $n,p,\rho$.

We can specify our analysis of quadratic fluctuation to the fluid. Consider a background $\phi^I=\alpha x^I$,
where $\alpha$ parameterizes the level of compression $B=\alpha^6$, and let us parameterize the perturbations by $\phi^I=\alpha (x^I+\pi^I)$.  
By expanding  eq.~(\ref{S_fluid}), integrating by parts and neglecting total derivatives
we get the Lagrangian for the Goldstones,
\be \label{L_sound}
{\cal L} = - F' (B) B\cdot (\dot \pi^I) ^2 + (F'(B)B + 2 F''(B)B^2) \cdot (\d_J \pi^J)^2   \; .
\ee
Only the longitudinal Goldstone has a gradient energy term; as
expected it is the only propagating degree of freedom:
sound waves are longitudinal.  The dispersion relation corresponds to a speed of sound $c_S^2$
\be
c^2_S  = \frac {F' + 2 F'' B\, B}{F'}\equiv  \frac{p'(B)}{\rho'(B)}= \frac{dp}{d\rho}\Big\vert _{\sigma= {\rm const}}\; ,
\ee
where in the last equality we have made use of the equations of state eq.~(\ref{rhop}) and have thus made contact with the usual expression for the speed of adiabatic sound waves in a fluid.
On the other hand, the transverse phonons, $\pi^i\propto \epsilon^{ijk}\partial_j f_k$ do not propagate. Their general solution at the infinitesimal level is linear in time
\be
\pi^i_T =\epsilon^{ijk}\partial_j\left (f_k(x)+g_k(x)t\right )
\ee
and corresponds to vortices in constant rotation. In the linearized approximation the fluid 4-velocity is indeed
$u^\mu=(1,-\dot \pi^i)$, so that we have\footnote{For large $t$ the linearized approximation for this solution breaks
down. Notice however that the bulk of the value of $\pi_T^i$ at large $t$ can
be eliminated by a volume preserving diffeomorphism around the stationary
solution. This means that what matters physically is just the time derivative
$\dot \pi_T^i$, which is just the fluid local velocity.}
$v^i=\epsilon^{ijk}\partial_j g_k(x)$.  The trivial time evolution of
infinitesimal vortex perturbations follows directly from the internal volume
preserving diffeomorphism invariance of the fluid, eq.~(\ref{noshape}).
% \footnote{This symmetry forces the Lagrangian to depend just on the determinant $B$, thus implying eq.~(\ref{L_sound}).}
 As a matter of fact, (also at the full non-linear level) this symmetry
implies the existence of an infinite set of conserved currents and of the
associated charges 
\be J^f_\mu =F'(B) B \, B_{IJ}^{-1} \, \partial_\mu\phi^J
f^I(\phi) 
\ee 
where $f^I$ is any vector function satisfying $\partial
f^I/\partial\phi^I=0$. The conservation of the infinite set of associated
charges basically corresponds to the trivial dynamical evolution of the
vortices.

This can be more explicitly seen by working in the comoving coordinate frame,
{\it i.e.}~by choosing the $\phi^I$ as space coordinates and by taking
$x^i(\phi,t)$ as the dynamical variable. From the basic relation
$\phi^I(x(\phi,t),t)={\rm const}$ we obtain by differentiation
\be
\dot\phi^I=-\partial_k\phi^I\dot x^k\equiv -W^I {}_k \, \dot x^k\, .  
\ee
 By
noticing that ${\rm det} (\delta^{ij}-\dot x^i\dot x^j) = (1-{\dot x}^2)$ we
have \be B= ({\rm det} W)^2(1-{\dot x}^2) \; , \ee so that the action now
reads \be
\label{newaction}
S = \int dt \,d^3\phi\,\frac{F\left(({\rm det} W)^2(1-{\dot x}^2)\right
)}{\vert{\rm det} W\vert}\, .  \ee This action is manifestly invariant under
volume preserving diffeomorphisms acting on the comoving coordinates $\phi^I$.
On the under hand Lorentz symmetry acts now as a combined transformation of
the time variable $t$ and the dynamical fields $x^i(\phi,t)$, and invariance
of the action is not explicitly manifest.  The volume preserving diffs are
realized on the dynamical variables as $\delta x^i = f^I\partial_I x^i$, with
$\partial_I f^I=0$, so that the set of conserved charges is given by \be
\label{charges}
C^f=\int d^3\phi \frac{\delta L}{\delta {\dot x}^i}f^I\partial_I x^i \ee 
For
$f^I=a^I={\rm const}$ the 3 conserved charges just correspond to spatial
momentum. The genuine fluid invariants are obtained by choosing
$f^I=\epsilon^{IJK}\partial_J \big(a_K\delta^3(\phi-\phi_0)\big)$, for
arbitrary constant $a_K$ and $\phi_0$. By eqs.~(\ref{newaction},
\ref{charges}) we obtain, after integration by parts, that the local 2-form
\be \partial_J\Big [ F' |{\rm det} W| \, {\dot x}_i\partial_K x^i\Bigr
]d\phi^J\wedge d\phi^K=d(F' {\sqrt B} \, u_i d x^i) \ee is time independent,
where in the last equality we have used the relation ${\dot x}_i=\sqrt
{1-{\dot x}^2} \, u_i$ for the space components of the 4-velocity. Considering
any 2-surface $\Sigma$ bounded by a closed circuit $C=\partial \Sigma$ we have
then \be
\label{kelvin}
\frac{d}{dt} \, \oint_C F'{\sqrt B}\, u_i dx^i \, =\frac{d}{dt} \, \int_\Sigma d\left [F'{\sqrt B} \, u_i dx^i \right ]\, =\, 0\,.
\ee
From the point of view of coordinate space the circuit $C$ depends on time as it moves with
the fluid flux, so the statement is that  the line integral of $F'{\sqrt B}u_i$ is conserved along the flux.
This quantity computed for all closed paths $C$ measures indeed the vorticity of the fluid's configuration.
Our result states that vorticity, thus defined, does not evolve with time.
This result corresponds to the relativistic generalization of Kelvin's theorem. In the non-relativistic limit,
we have $p\ll \rho$, which from eq.~(\ref{rhop}) corresponds to $F(B)\simeq \sqrt B$. In this case
the conserved quantity becomes simply the circuit integral of the velocity $v_i$, as stated in Kelvin's theorem.

In the last discussion Lorentz invariance was not explicitly manifest. We can however easily express
the result in eq.~(\ref{kelvin}) in a manifestly Lorentz covariant manner. From the definition of
the 4-velocity and from eq.~(\ref{kelvin}) the relevant one-form we should focus on is
\be \label{starA}
F'{\sqrt B} \, u_\mu dx^\mu  \equiv V_\mu d x^\mu \,.
\ee 
Using $V$ the equations of motion
\be
\partial_\mu \left [F' B \, B^{-1}_{IJ}  \, \partial^\mu \phi^J\right] = 0 
\ee
are written conveniently as
\be
\label{eomform}
(d  V) \wedge d\phi^J \wedge d\phi^K \epsilon_{IJK}=0
\ee
The geometric implication of this formula is that $(d V)=0$ on any two surface $\Sigma$
 tangent  at all points to the four velocity $u^\mu$. This is easily seen by 
working in comoving coordinates, where $u^\mu=(1,0,0,0)$ and eq.~(\ref{eomform})
reads $(d V)_{0I}=0$ for $I=1,2,3$. 
We can create such a surface $\Sigma$
by choosing a spacelike closed cycle $C_0$ and by considering the world-surface it describes along the fluid motion up to some final cycle $C_1$.  We have $\partial \Sigma = C_0-C_1$ so that by Stokes theorem we find
\be
\label{covariantvorticity}
0=\oint_\Sigma d V= \oint_{C_0} V-\oint_{C_1} V
 \ee
As expected, the quantity $\oint \! F'{\sqrt B} \, u_\mu dx^\mu$ is conserved along the flow. 
Notice however that  neither $C_0$ nor $C_1$ has to be necessarily chosen at fixed time.
To summarize, our result states that given a two-surface $\Sigma$ with the topology of a cylinder and tangent to the local 4-velocity at any point, the integral of $V = F'{\sqrt B} \, u_\mu dx^\mu$ around $\Sigma$ is independent of the cycle chosen to compute the integral.

\subsection{From Fluid to Superfluid and to Ghost Condensate}
\label{ghostcond}
We have seen that vorticity is conserved through the time evolution of the system, corresponding to the `trivial' dynamics of the vortices. In particular if vorticity vanishes on any initial time surface it will vanish
throughout the time evolution. For this subclass of the fluids configuration it is natural to expect
our system to be described by just one scalar field instead of three. Let us see how this works.

The most direct, although at first sight  unintuitive, way to proceed is to introduce
an auxiliary 1-form $V_\mu$  and start with the following action
\be
\label{Gaction}
{\cal S}_{V,\Omega} =\int G(X) -V\wedge \Omega
\ee
where, $X=-V^\mu V_\mu$ and $\Omega$ is the fluid volume form defined before. 
The $V$ equations
of motion
\be
\label{Gprime}
-2 G'(X)V^\mu =(\star \Omega)^\mu \equiv J^\mu 
\ee
allows us to algebraically solve for $V_\mu$ as a function of $J^\mu$ and to write
eq.~(\ref{Gaction}) as a functional of the fluid variables 
\be
\label{legendre}
\int G(X) -V\wedge \Omega\Big \vert_{\delta S/\delta V=0} = \int F(B)\, .
\ee
The functions $G(X=-V^2)$ and $F(B=-J^2)$ are the Legendre transforms of one another, and $V_\mu$ and $J^\mu$ are the conjugated variables. By the usual properties of Legendre transforms, eqs.~(\ref{Gaction}, \ref{legendre}) also imply the conjugate to eq.~(\ref{Gprime}) 
\be
\label{Fprime}
2 F'(B)J_\mu=V_\mu\, .
\ee
Up to a factor of 2, the form $V$ coincides on-shell  with the form $V$ we defined in the previous section, hence our notation. (This is  also evident by varying eq.~(\ref{Gaction}) with respect to $\phi_I$ and comparing to eq.~(\ref{eomform})). Notice that eqs.~(\ref{Gprime}, \ref{Fprime}) imply the scalar
relation
\be
\label{dual}
2 F'(B) =\frac{-1}{2 G'(X)}
\ee
as well as 
\be
\label{BvsX}
B=X (G'(X))^2\qquad X=B(F'(B))^2\, .
\ee
 Moreover eq.~(\ref{legendre}) can be also
written as 
\be
\label{gminusf}
G-F=-2BF'=2XG'\, .
\ee

We have now all the ingredients to discuss fluid motion at zero vorticity.
From the discussion at the end of last section, this situation is realized when
\be
\label{potential}
V = d \psi
\ee
where $\psi$ is a scalar field. In this case $u_\mu\propto \partial_\mu \psi$, corresponding 
to the Lorentz covariant generalization of potential flow in a fluid \cite{LL}. When
eq.~(\ref{potential}) holds, the equation of motion for $\phi$ are trivially satisfied. However the dynamics of $\psi$ is non-trivial and is determined by taking the divergence of eq.~(\ref{Gprime})
\be
\partial_\mu (G'(X) \partial^\mu \psi)=0\, .
\ee
We have therefore established that the solutions in the absence of vorticity are into one-to-one
correspondence with the solutions for  the dynamics of a derivatively coupled scalar with Lagrangian
$L=G(-\partial_\mu \psi \partial^\mu \psi)$. Moreover since $\partial_\mu \psi \propto u_\mu$ is a time-like  directed
vector, the scalar $\psi$ depends linearly on time in the ground state  $\psi = c \, t$. 
This is precisely the dependence of the charged field phase in a superfluid. We have thus 
established that, in the absence of vortices, the classical dynamics of a fluid 
is equivalent to that of a superfluid with suitably chosen Lagrangian (cf. \cite{Son:2002zn}).
As a further check of this equivalence, notice that the energy momentum tensors in the 
two descriptions
coincide. This is manifest by eq.~(\ref{legendre}), since the fluid and superfluid action differ
by the integral of the four form $V\wedge \Omega $ which does not depend on the metric.
It can also be checked directly by using eq.~(\ref{gminusf}) \footnote{Indeed we originally derived
our relation between fluid and supefluid by equating their energy-momentum and 
number current. This leads precisely to the relation described at the beginning of this section.
We believe however the Lagrangian approach we followed is analitycally more transparent.}.

The relation between fluid and superfluid is a close relative of the duality
between a 2- and a 0-form.  Consider indeed eq.~(\ref{Gaction}), with $\Omega
= d A_2$.  The 2-form $A_2$ acts like a Lagrange multiplier setting $dV=0 $
thus reducing to a 0-form $V=d\psi$ action.  On the other hand, by solving
the $V$ equations first we would get a fully equivalent local action for
$A_2$. The fluid is distinguished from the two form because
$\Omega=d\phi^1\wedge d\phi^2\wedge d\phi^3$ is not the most general field
strength a two form can have (it is a function of 3 field variables, instead
of 6). However although the off-shell field space of the fluid is of smaller
dimensionality that for the 2-form theory, the space of solutions is
nonetheless bigger. This is perhaps counterintuitive, but it becames less so
if one recalls that in a gauge theory some of the field variables act just as
Lagrange multipliers. We do not want to explore these issues further here, but
just make a few extra remarks elucidating the duality character of the
fluid/superfluid relation. Notice that dynamical equations and structural
constraints are interchanged. So the conservation of $J_\mu$ is structural for
the fluid while it follows from the equations of motion in the
supefluid. Conversely vorticity is dynamically conserved in the fluid and
structurally zero in the superfluid. More importantly the fluid/superfluid
relation is only valid for the classical action on-shell, and breaks down when
quantum corrections are taken into account. In fact in a duality relationship
the quantum perturbation series are not mapped into one another, and when one
description is weakly coupled that other is not.

This last property and others are made clear by focussing on a particular
superfluid, the so called ghost-condensate model, which also provides an
interesting and consistent modification of gravity \cite{Arkani-Hamed:2003uy}.
The model is defined by assuming that the function $G$ is positive and has a
minimum (where $G'=0$) at some finite point $X=c_*^2$ in field space. This is the
ghost-condensate point.  At this point, although the field is time dependent
$\psi =c_* t$, the energy momentum tensor is still proportional to the
metric. Thus the universe is in a de Sitter phase with an effective
cosmological constant $\equiv -G(c_*^2)$, which differs from the genuine
cosmological constant defined at zero field $-G(0)$. What does this point
correspond to in the fluid description?  By the first of eqs.~(\ref{BvsX}) the
answer is $B=0$, i.e the point of infinite dilution, while, not surprisingly,
eq.~(\ref{dual}) tells us that this is a singular point. More precisely, by
solving the second of eqs.~(\ref{BvsX}) we find two solutions, corresponding
to a branch singularity \be F_\pm(B)=-{\sqrt B}f_\pm(B)+ F_0 \ee where $f_\pm$
are smooth functions satisfying \be \lim_{B\to 0} f_\pm(B) = \pm c_*\, .  \ee
$F_+$ and $F_-$ respectively corresponds to the superfluid branches $X>c_*^2$
($G'>0$) and $X<c_*^2$ ($G'<0$). As $B\to 0$ the $F_+$ branch becomes a
diluted dust with positive energy density $\rho \sim c_* \sqrt B$ and positive
pressure\footnote{By eq.~(\ref{gminusf}) we have $p=B^{3/2} f'_+\equiv
G(X)-G(c_*^2)\geq 0$.} superposed to a cosmological constant
$F_+(0)=F_0$. Notice that by eq.~(\ref{gminusf}) $F_0=G(c_*^2)$, which propels
de Sitter expansion, is now truly the cosmological constant. On the ther hand,
the branch $F_-$ corresponds to a diluted fluid with negative energy density
$\rho \sim -c_* \sqrt B$ but positive pressure $p=B^{3/2} f'_-(B)$.  The two
branches $X>c_*^2$ and $X<c_*^2$ thus correspond to two distinct fluids. These
two fluids cannot mix as they are always separated by a region of infinite
dilution, where $B=0$. Moreover the regions corresponding to $F_+$ will tend
to expand as they correspond to regions of higher density and pressure in an
ordinary fluid. The regions corresponding to $F_-$ will instead contract~\cite{Krotov:2004if,Arkani-Hamed:2005gu} as
they correspond to a fluid with negative energy density but whose pressure
grows with compression\footnote{This follows from Euler equations where
acceleration along fluid moton is proportional to $\nabla p/\rho$.}. Notice
also that in the region with $G'<0$ the superfluid theory features a gradient
instability and this is consistent with the imaginary speed of sound implied
by eq.~(\ref{L_sound}). However in the fluid description, unlike for the
superfluid, the corresponding mode is also a ghost, in the sense that its time
derivative term has the wrong sign in the action, see
eqs.~(\ref{L_sound}, \ref{dual}).  This makes sense since duality corresponds
to a canonical transformation that exchanges $p\to-q$ and $q\to p$ for some of
the variables\footnote{ Example $B$ and $E$ in electric-magnetic duality.}, so
that if in one description the potential is negative in the dual description
it is the kinetic term to be negative.  As a final remark, the quantum cut-off
$\Lambda$ of the fluid theory goes to zero like $\rho^{1/4}\sim B^{1/8}$ as
$B\to 0$. Therefore the fluid is formally infinitely strongly coupled at the
ghost-condensate point\footnote{For ordinary fluids associated to weakly
coupled theories, the physical cut-off $\Lambda_{\rm phys}$ where the hydrodynamic
description breaks down is always smaller than $\rho^{1/4}$ and is determined
by the inverse mean free path. For instance in a hot plasma we have
$\Lambda_{\rm phys}\sim g^2 T\ll T\sim \rho^{1/4}$.}  while the superfluid has a
finite UV cut-off, and thus represents a well to do effective field theory.
This, and the presence of two distinct phases with opposite energy density,
represent the novelty of the ghost condensate model with respect to ordinary
fluids.

\section{Degenerate cosmic supersolids}

\label{symmetries}

Throughout the paper we have been focussing on the scalar field excitations,
neglecting the backreaction of gravity.  As we explained in the beginning,
this is consistent as long as we limit ourselves to distances $r\ll r_G=
M_{Pl}/\Lambda^2$, where $\Lambda$ is the characteristic energy scale of the scalar Lagrangian. 
The gravitational length $r_G$ normally controls both the
mixing between the Goldstone boson and the graviton, basically the graviton
mass, and the curvature of space time\footnote{We are treating the function
$\tilde F$ (see eq.~(\ref{parametric})), as well as its derivatives and $\partial \phi/\Lambda^2$ as $O(1)$
quantities. }. Because of the coincidence of these two length scales, the
mixing between graviton and Goldstones cannot normally be discussed around
flat space, and is thus qualitatively different from the Higgs phenomenon in
ordinary gauge theories. One can however think of peculiar situations where
$|T_{\mu\nu}|\ll \Lambda^4$ can be practically neglected, while the graviton
mass is still of order $1/r_G$. In this situation, the propagation of gravity
is modified at distances $r_G$ where curvature can be neglected, thus defining
infrared modifications of gravity that are the analogue of the ordinary Higgs
mechanism. These theories include the Fierz-Pauli model, the ghost condensate
model \cite{Arkani-Hamed:2003uy} and the general class of massive gravities of
\cite{Dubovsky:2004sg} and are indeed particular cases of the class of scalar
theories  that we have been focussing on.  In
these models, the difficulties and peculiarities associated to the dispersion
relation of the Goldstone fields are also directly connected to the difficulty
in violating NEC in general.  Indeed when $T_{\mu\nu}=0$ eq.~(\ref{chain})
simply reduces to $ L(p,q)-L(q,p)=0$. Since the matrix $A^I_\mu$ is non-zero,
it immediately follows that either $\Gamma_-$ or $\Gamma_+$, or both, do not
exist. Therefore, these theories can only be acceptable in limiting situations
in which some dispersion relation is degenerate, {\it i.e.}~either $\omega=0$
or $k=0$, and the corresponding mode does not propagate. This is in agreement
with the results of Ref.~\cite{Dubovsky:2004sg} where the conditions for
stability were studied in detail. Because of these degenerate dispersion
relations, the stability of these models is threatened both by higher order
effects, ex. higher derivative terms, and by slight deformations of the
background.
%Consistently with our present general analysis, it was found that in generic cases there are instabilities, either tachyonic-like, or ghost-like. 
%However, for some very specific forms of the function $F$,  it was found that there can be some stable situations. In every such cases, there are some non-propagating modes that have dispersion relation of the form $p=0$ or $\omega=0$.  
However in Ref.~\cite{Dubovsky:2004sg} it was also shown that in some cases
the singular dispersion relations can be protected by internal symmetries of
the scalar manifold, thus ensuring in a natural way the stability of the
system.

Now, in the presence of these non-propagating modes, the general discussion of
the stability conditions in section~\ref{proof} does not apply directly, since
there we had assumed that the determinant of the kinetic matrix
$L_{IJ}(p)/k^2$ had $2N$ single roots. Therefore, as in
\cite{Dubovsky:2004sg}, we must take a case-by-case approach, and do a very
similar analysis to what was done there, to look at what happen if we relax
the assumption that the energy-momentum tensor vanish. In particular we would
like to know if there are stable cases when $w < -1$.
  
The case we study is not as general as the one considered in section $4$ as we
consider only $4$ fields $\phi_I$ with $I=0,1,2,3$ in a rotationally invariant
situation. However, the assumption of rotational symmetry is not a limitation
when thinking of cosmological applications, while the 4-fields corresponds to
the simplest case where the full translation group is broken.  
Setting as before $\Lambda=1$ the background
values for $\phi_I$ are taken to be
\begin{eqnarray}
\label{background}
\phi_0 &=& \alpha t  \\
\nonumber \phi_i &=& \beta x^i \, \, \; i=1,2,3 \,.
\end{eqnarray}
Here, $\alpha$ and $\beta$ are arbitrary. By the discussion in section \ref{fluids} this system consists of a superfluid ($\phi_0\propto t$) interacting with a solid ($\phi_i\propto x^i$), hence the denomination supersolid.

The Lagrangian for the fluctuations $\pi_0, \pi_i$ around the background
(\ref{background}) can be written, in the notation we have been using so far,
as
\begin{multline}
\left(F_{00} + 2  F_{00,00} \alpha^2\right) \left(\partial_t \pi_0 \right)^2 + \left(-F_{00} +  F_{0i,0i} \frac{\beta^2}{2} \right) \left(\partial_i \pi^0\right)^2 + \left(\frac{ F_{0i,0i}}{2} \alpha^2 + \frac{F_{ii}}{3} \right) \left(\partial_t \pi_i \right)^2 \\
- \frac{1}{3}  \left(4 F_{00,ii} + F_{0i,0i} \right) \alpha \beta \pi_0 \partial_j \partial_t \pi_j + \left(-\frac{1}{3}F_{ii} +  \beta^2 \frac{3 F_{ii,jj} -F_{ij,ij}}{48} \right) \left(\partial_i \pi_j \right)^2 \\ +  \beta^2 \left(\frac{3 F_{ij,ij} - F_{ii,jj}}{48} \right) \left(\partial_i \pi_i \right)^2
\end{multline}
We can also calculate the energy-momentum tensor in this background,
\begin{equation}
T_{\mu \nu} = c \eta_{\mu \nu} + \left(\alpha^2 F_{00} \delta_{\mu 0} \delta_{\nu 0} + \frac{1}{3} F_{ii} \beta^2 \delta_{\mu j} \delta_{\nu j} \right) 
\end{equation}
The null energy condition is violated if
\begin{equation}
p + \rho = \alpha^2 F_{00} + \frac{1}{3} F_{ii} \beta^2 <0 
\end{equation}
To analyze this system, it is convenient  to rescale $\pi_0$ and $\pi_1$ and rewrite the Lagrangian  as 
in \cite{Rubakov:2004eb,Dubovsky:2004sg},
\begin{multline}
\label{Lsergei}
2 m_0^2 \left(\partial_t \pi_0 \right)^2 + m_1^2 \left(\partial_t \pi_i \right)^2 + \left[m_1^2 - (p+\rho) \right] \left(\partial_i \pi_0 \right)^2 + \left( 4 m_4^2 - 2 m_1^2\right) \pi_0 \partial_t \partial_i \pi_i  \\
- m_2^2 \left(\partial_i \pi_j \right)^2 - \left(m_2^2- 2 m_3^2 \right) \left(\partial_i \pi_i \right)^2
\end{multline}
The only difference with the Lagrangian of \cite{Dubovsky:2004sg} is the appearance of $p+\rho $ in the coefficient of $(\partial_i \pi_0)^2$. 
As we explained before, for rotational invariant systems the positivity of the Hamiltonian is a necessary and sufficient condition for stability.
In eq.~(\ref{Lsergei}) for $p+\rho<0$, the gradient energy is made positive by having $m_1^2$  sufficiently large and negative. In this case, however, the coefficient of   $(\partial_t \pi_i)^2$ would have the wrong sign, and the Hamiltonian
would still be not positive definite. We thus recover our general result that, in the rotational invariant case, we cannot achieve stability with all the modes propagating.
We have thus to consider the degenerate cases.
The analysis  is immediate in the vector sector 
(the modes $\pi_i^T$ that satisfy $\partial^i \pi_i^T = 0$) 
where the Lagrangian is simply
%the same in Ref.~\cite{Dubovsky:2004sg}
\begin{equation}
\mathcal{L}_{\text{vector}} = m_1^2 \left(\partial_0 \pi_i^T \right)^2 - m_2^2 \left(\partial_i \pi_j^T \right)^2
\end{equation}
 In the scalar sector, consisting of the modes $\pi_0$ and $\pi_L$ defined by $\pi_i = 1/\sqrt{\partial_i^2} \partial_i \pi_L$, the Lagrangian can be written in momentum space as
 \begin{equation}
 \label{eq:momentumspace}
k^2 \begin{pmatrix} \pi_0 & \pi_L \end{pmatrix} \begin{pmatrix} m_0^2 \nu^2 +m_1^2 - (p + \rho)& -i \mu \nu \\  i \mu \nu & m_1^2\nu^2-\eta\end{pmatrix} \begin{pmatrix} \pi_0 \\ \pi_L \end{pmatrix} \, ,
\end{equation}
 with $\nu = \omega/k$, $\mu = (2m_4^2-m_1^2)$ and $ \eta = 2(m_2^2-m_3^2)$.
 The analysis of the stability of this theory is very similar to the analysis done in \cite{Dubovsky:2004sg}. The absence of ghosts and gradient instabilities constrains the sign of the coefficients in the diagonal entries of the kinetic matrix
 \be
 \label{constraints}
 m_0^2>0\, \qquad m_1^2>0\,,\qquad -m_1^2 + (p + \rho) >0\, ,
 \qquad \eta>0 \ee This condition agrees with our general result: for
 $p+\rho<0$ there cannot be stability if all modes are normal. We must however
 analyze with care the situation where some of the modes become degenerate,
 corresponding to some root $\nu$ going to either $0$ or $\infty$.  If the
 unstable modes can be forced to these limiting dispersion relation by
 consistent symmetry requirements, then the system has a chance of becoming
 acceptable. The limiting situations arise when one or more of the
 coefficients in eq.~(\ref{constraints}) equal 0. For instance for $m_0^2=0$
 one of the roots is $\nu=\infty$ and the corresponding field has a pure
 gradient quadratic action. By the arguments in
 refs. \cite{Rubakov:2004eb,Dubovsky:2004sg}, we may thus forget about this
 mode, provided it does not reappear when higher derivative terms are
 considered or the solution is slightly perturbed. The other mode has a normal
 dispersion relation and is not a ghost as long as $m_1^2,\eta >0$,
 irrespective of $p+\rho<0$.  However we know of no symmetry that can
 naturally enforce $m_0^2=0$ without introducing extra unprotected degenerate
 modes, so this possibility does not seem viable.  Similarly, the case
 $\eta=2(m_2^2-m_3^2)=0$ allows to send the unwanted root at $\omega=0$ while
 keeping $p+\rho<0$.  But also this second choice cannot be enforced by
 symmetries.
 
%   From the general 
% Similarly to what was found in that paper, we find that for general $m_0,m_1,m_2,m_3$ and $m_4$ the Lagrangian (\ref{eq:momentumspace}) has classical instabilities or ghosts. 

%Therefore, we are forced to some corner of parameter space, where some modes are not propagating. 

In Ref.~\cite{Dubovsky:2004sg} several internal symmetries of the scalar field
manifold which protect the required degeneracies at $T_{\mu\nu}=0$ were presented. These symmetries can be thought of as
generalizations of the volume preserving diffeomorphism invariance which is
relevant for ordinary fluids.  In what follows we would like to survey the
various symmetries to see whether they can make the $p+\rho<0$ case
acceptable.
%
%Let's begin by studying the cases where the dispersion relations for these modes (either $\omega =0$ or $p=0$) are protected by a residual subgroup of diffeomorphism invariance. 

First we consider the symmetry
\begin{equation}
\label{supersolid}
\phi^i \rightarrow \phi^i + \xi^i(\phi^0) \;,
\end{equation} 
which forces $m_1=0$. This symmetry implies the presence of three
instantaneous sound waves with dispersion relation $k=0$, so we may call this
system a rigid supersolid.  Notice that while for a fluid the internal
symmetry (see eq.~(\ref{noshape})), is making the solid totally deformable,
eq.~(\ref{supersolid}) is making the solid infinitely rigid. 
 Cosmological and phenomenological properties of rigid supersolids
with zero energy-momentum were studied in
\cite{Dubovsky:2005dw,Dubovsky:2004ud}.  In general symmetry
(\ref{supersolid}) implies that the vector modes have a dispersion relation of
the form $k=0$.  The Lagrangian for the scalar modes can be written as
\begin{equation}
k^2 \begin{pmatrix} \pi_0 & \pi_L \end{pmatrix} \begin{pmatrix} 2 m_0^2 -
 (p + \rho) & -2 i m_4^2 \nu \\ 2 i m_4^2 \nu & 2 (m_3^2-m_2^2) 
\end{pmatrix} \begin{pmatrix} \pi_0 \\ \pi_L \end{pmatrix} \, ,
\end{equation}
According to the previous discussion one mode has dispersion relation $k=0$
 and the second one
\begin{equation}
\omega^2 = \frac{p + \rho}{ \left(m_0^2 - \frac{m_4^4}{m_3^2-m_2^2}\right)} k^2\;.
\end{equation}
For real $\omega$ the sign of the corresponding residue is the same of $p+\rho$ so that this mode is a ghost when NEC is violated.
%If $p+\rho <0$, then in order for $\omega_0$ to be real and avoid classical instabilities, we need $m_0^2 < m_4^4/(m_3^2-m_2^2)$.  However, it is easy to show that this implies 
%\begin{equation}
%\left.\frac{d \lambda}{d \nu} \right|_{\nu_0} < 0 \, ,
%\end{equation}
%where $\lambda(\nu)$ is the eigenvalue that has a zero at $\nu_0$, indicating the presence of a ghost. Note that if $p + \rho = 0$, higher derivative term must be included to get a dispersion relation of the form $\omega^2 \propto p^4$. 

Another interesting symmetry that works in case of massive gravity is
 \be
 \phi^i \rightarrow \phi^i(\phi^j)
 \ee
that is the full group of diffeomorphisms of the $\phi^i$ submanifold.
Notice that, although this is a very big symmetry, one can still write a non-trivial action 
thanks to the presence of $\phi^0$. 
This symmetry implies $m_2=m_3=m_4=0$, and the presence of three sound waves with
zero propagation velocity, $\omega=0$. Consequently, a system possessing such a symmetry may be called a
superdust. Both  vector modes in the superdust have dispersion relation $\omega=0$. 
The scalar sector is characterized by the following kinetic term,
\begin{equation}
k^2 \begin{pmatrix} \pi_0 & \pi_L \end{pmatrix} \begin{pmatrix} 2 m_0^2 \nu^2 +m_1^2- (p + \rho) & i m_1^2 \nu \\ - i m_1^2 \nu & m_1^2 \nu^2 \end{pmatrix} \begin{pmatrix} \pi_0 \\ \pi_L \end{pmatrix} \, ,
\end{equation}
The eigenvalues of the kinetic matrix have zeros at $\omega=0$ and,
 $\omega^2= k^2 \, (p+\rho)/2  m_0^2$. 
Again, in the case of $p+\rho <0$, gradient 
instabilities can be avoided by taking $m_0^2<0$, but then the propagating mode is a ghost. 
Similarly, the symmetries $\phi^0 \rightarrow \phi^0 + \xi^0(\phi^0)$ 
and $\phi^0 \rightarrow \phi^0+ \xi^0(\phi^i)$ do not lead to stable situations.
%
%In other phases of this theory, the analysis of the stability of the theory is almost identical to what was done in \cite{Dubovsky:2004sg}. For example, in the  $m_0^2=0$ phase, we find a stable situation for $\beta>0$ and $\mu^2 m_1^2 <m1^2 - (p+\rho)/(\Lambda^2 )$, and for $m_2^2=m_3^2$, we need $\alpha>0$ and $\mu^2 -1+(p+\rho)/(m_1^2 \Lambda^2) >0$ to avoid instabilities. However, these choices of parameters are not protected by any symmetries.

We conclude that for $p+\rho<0$ not even the extra symmetries that
helped make sense of modifications of gravity can help avoid 
instability. $p+\rho=0$ emerges once again as a truly limiting 
case even for this class of generalized cosmic solids.

\section{Towards $w<-1$}
\label{towards}
One of the motivations to study NEC violations 
in  effective field theories is to 
understand whether the deceleration parameter of the cosmological
 expansion can be strongly negative, $w<-1$. It is then natural
to ask  what are in this respect the implications of our results. Is there still room for a cosmology with increasing\footnote{Notice that there are
proposals~\cite{Lue:2004za,Csaki:2005vq} that mimic the presence of a
 dark energy component with $w<-1$ while
the Hubble parameter still always decreases with time. In other words
in these models the total $(p+\rho)$, dark energy plus
dark matter, is still positive. } Hubble parameter?

Our results make it very unlikely that such a cosmology may be driven by some
NEC violating field theory.  However one should keep in mind that our
considerations, strictly speaking, apply only to scalar theories in the
leading derivative approximation. For instance we did not consider situations
where higher derivative terms in the effective action are important. An
example of a non-trivial higher derivative theory from which to start this
study could be offered by the dynamics of the brane bending mode in the DGP
model \cite{Dvali:2000hr, Nicolis:2004qq}.  Moreover our analysis of the
symmetries which could protect the degenerate dispersion relations did not
cover the most general case, even for scalar theories.  Also it may be
possible to construct models where instabilities are present, but slow enough
to be benign.  On the other hand, we expect our results to be quite robust,
especially because they relate NEC violation to the instabilitites in a sector
of the theory, the Goldstones of spontaneously broken space-time translations,
that exists whenever there is a non-trivial field background.

Here we would like to discuss yet another possibility for having $w<-1$.
The basic idea is that the
accelerated expansion of the Universe is  due to a genuine ``modification
of gravity'', by which the observed metric depends on new degrees of freedom
and thus does not satisfy
 %in other words the observed metric may not satisfy 
 the Friedmann
equation. For instance, one may imagine a probe 3-brane moving in a curved
bulk space. This brane may be accelerated by  forces of 
non-gravitational origin (e.g., it may be charged under some form fields).
Then the induced metric on the brane will evolve in time even in the absence
of an energy-momentum tensor localized on the brane. Clearly, the challenge for
 proposals of this type is to explain why gravity is  described by the 4-dimensional
Einstein theory at short length scales. Scalar tensor
theories of gravity are also simple enough arena to toy with this idea. This was indeed done in 
Ref.~\cite{Carroll:2003st}  (for earlier work see \cite{Boisseau:2000pr}) with the conclusion that ``only highly contrived
models would lead observers to measure $w<-1$''. So our first
purpose here
is to understand in simple terms the basic reason for contrivedness in such models.
This will allow us to propose a relatively simple candidate model where 
$w<-1$ can be achieved
without fine-tuning\footnote{More precisely,  in the model we shall present there is no extra
fine-tuning associated to $w<-1$. Still there is a necessity
to fine-tune the potential of  certain scalar field to make it very smooth. This is
a consequence of the presence of a dynamical ``dark energy'' with
$w\neq -1$. This type of fine-tuning is inherent in most ``quintessence''
scenarios.}.  A minimal 4d setup possessing all the ingredients needed to
implement this idea is described by the following action \be
\label{setup1}
S[g,\psi,\phi]=
M_{Pl}^2\int d^4x\sqrt{-g} R+S_m[\psi,f(\phi/M)g_{\mu\nu}]+
S_\phi[\phi,g_{\mu\nu}] \; .
\ee
Here $\psi$ collectively denotes the Standard Model fields, while $\phi$ is
an extra scalar field.  The goal is to construct an action of the form 
(\ref{setup1}) allowing for a solution with the following properties.

{\bf (A)} The field $\phi$ changes in time and the
 physical metric $f(\phi)g_{\mu\nu}$ has the FRW form. 
The main source
of visible FRW expansion is  the time dependence of the factor $f(\phi/M)$. In particular, this implies
that the energy density of the field $\phi$ itself should be small enough not to cause significant 
backreaction.

{\bf (B)}
Deviations from  Einstein theory at  short length scales due to  
coupling of the field $\phi$ to the matter 
energy-momentum tensor are small enough.

In other words, the field $\phi$ is a non-gravitating cosmic  clock.
Let us see that a conventional scalar field with the action
\[
S_\phi=\int d^4x\sqrt{-g} (\d_\mu\phi)^2-V(\phi)
\]
is very unlikely to play such a role. 
In order to proceed, let us first assume that the interaction of $\phi$ with
matter is weaker than gravity, $M\gg M_{Pl}$.
In this case, condition {\bf (B)} is satisfied and the time evolution of the 
$f(\phi/M)$ factor in the physical metric gives a contribution to the observed
expansion rate of the Universe 
\be
\label{Hf}
H_f^2=\l{\dot f\over f^2}\r^2\sim {\dot\phi^2\over M^2}\;,
\ee
where we assumed that $f(x)$ is a generic smooth function of order one.
On the other hand there is a contribution to the expansion rate of the 
Universe from the energy-momentum tensor of the field $\phi$ itself, which
is at least of order
\be
\label{Hphi}
H_\phi^2\sim {\dot\phi^2\over M_{Pl}^2}\;.
\ee
The latter contribution gives $w>-1$ and dominates the first one in the 
considered regime, $M\gg M_{Pl}$.

Let us now discuss the case in which the clock field $\phi$ is coupled to 
matter more strongly than gravity, $M\lesssim M_{Pl}$. In this case measurements
of the deflection of light by the Sun combined with tests of the Newton law
at sub-millimeter scales imply that $\phi$ should have rather big mass,
\be
m_\phi\gtrsim \mbox{0.1 mm}^{-1}\;.
\ee
Now, during a Hubble time  $\phi$ varies by
\be
\Delta\phi\sim M\;
\ee
so that the potential $V(\phi)$ gives a contribution
 to the energy density of order 
\be
\rho_\phi\sim m_\phi^2\Delta\phi^2\sim m_\phi^2M^2\;.
\ee
In order for this energy to be subdominant in today's Universe one needs
\be
M\lesssim  m_\phi\cdot\l \mbox{0.1 mm}\r^{-2}\;.
\ee
In other words, the coupling of the clock field to matter becomes non-perturbative
below  $0.1$~mm. It is hard to imagine how such a
 theory could make sense.

To summarize, we see that in both cases the essence of the problem is
that the gravitational backreaction of the cosmic clock field $\phi$
tends to give a larger contribution to the expansion rate 
than the direct coupling $f(\phi)$.

One way to avoid this problem might perhaps be to consider the case 
$M\lesssim M_{Pl}$ and try to hide the field $\phi$ through the
chameleon mechanism \cite{Khoury:2003rn}, instead of introducing the large mass $m_\phi$.

Another possibility, instead,  is to consider a clock field coupled
to matter more weakly than gravity, $M\gg M_{Pl}$,  and to try to
minimize its gravitational backreaction. 
 A natural  candidate to  the role of non-gravitating  clock 
is the ghost condensate
\be
S_\phi[\phi,g_{\mu\nu}]=\int d^4x\sqrt{-g}\Lambda^4 G(X)
\ee
where 
$X=(\d_\mu\phi)^2/\Lambda^4$ as in sect.~\ref{ghostcond} but we also assume
\be
G'(X_0)=G(X_0)=0\;.
\ee
for some $X_0$.
Without loss of generality we may set $X_0=1$. Note that the direct coupling to
matter breaks the shift symmetry of the ghost condensate. Then,
in general, matter loops will generate a potential for the field $\phi$,
which is too large for the construction below to work. The fine-tuning needed
to cancel this potential is not related to the requirement $w<-1$ and
is similar to one present in most quintessence scenarios.
Then one finds the following ``cosmological''
solution,
\be
g_{\mu\nu}=\eta_{\mu\nu},\;\phi=\Lambda^2 t
\ee
with matter fields in their vacuum state. This solution provides an
extreme implementation of the original idea, in which the dependence
on the field $\phi$ is the only source of time evolution for the
physical metric. The Hubble parameter is given by (\ref{Hf})
and is of order 
\be
\label{H}
H_f\sim  \Lambda^2/M\;.
\ee 
Its time evolution can be fairly arbitrary, in
particular it can grow with time.
 To check that this
model may present interest for  cosmological applications let us check
that values of $H$ of the order of 
the current Hubble rate $H_0$ can be achieved
without contradicting to the tests of general relativity.
Two type of effects should be suppressed in this model. First there
is a modification of gravity due to the mixing between the metric and the ghost
condensate mode.  Second there is an extra scalar force arising from
the direct coupling of the trace of the energy-momentum tensor to the ghost
condensate field. Mixing with gravity gives rise to a Jeans-like instability
with  characteristic time-scale
\[
t_c\sim M_{Pl}^2/\Lambda^3\;.
\]
There are no sizable effects due to this mixing as long as $t_c$ is larger than
the current age of the Universe (and even for somewhat shorter values of 
$t_c$). This implies that as large values as $\Lambda\sim 100$~MeV are allowed.
Then eq.~(\ref{H}) implies that the parameter $M$ which determines the 
strength of the direct coupling of the ghost condensate field to matter can be
well above $M_{Pl}$ still providing a realistic value of $H$. The only
subtlety here is that in the limit in which gravity is decoupled the ghost condensate
has the  unusual dispersion relation
\be
\label{gcdisp}
\omega^2-{k^4\over\Lambda^2}=0\;,
\ee
with the $k^4$-term coming from higher-derivative terms.
Naively, this dispersion relation leads to the puzzling conclusion that the 
exchange of the ghost condensate field leads to a static potential which grows
proportionally to the distance from the source. This would imply the breakdown
of perturbation theory far from the source even for very large values of 
$M$. A resolution to this problem is that the dispersion relation (\ref{gcdisp})
implies that the velocity of the scalar quanta
\be
v={k\over\Lambda}
\ee
is very low for small values of $k$. As a result 
a very long time $\sim L^2 \Lambda$ is needed for this potential
to be established at a large  distance scale $L$. 
More precisely accounting for the mixing with gravity the
dispersion relation (\ref{gcdisp}) is modified to
\be
\omega^2-{k^4\over\Lambda^2}+{k^2\Lambda^2\over M_{Pl}^2}=0 \; .
\ee
The last term gives rise to the Jeans-like instability discussed above and 
dominates over the second one for
\be
k<m\equiv{\Lambda^2\over M_{Pl}} \; .
\ee
Consequently, for time-scales shorter than $t_c$ the linear potential
is established only for distances shorter than $m^{-1}$.  At larger distances
mixing with gravity trades this growing-with-distance-potential with an 
instability that develops in time. For small enough $\Lambda\lesssim 100$~MeV
and large enough
$M\gg M_{Pl}$ both effects can be neglected.

Note, that in the presence of  massive matter sources non-linear effects
dominate the dynamics of the ghost condensate~\cite{Arkani-Hamed:2005gu}. These effects may introduce
local spatial gradients of the field $\phi$ of order $\d_i\phi\sim \Lambda^2$.
Then direct coupling of the ghost condensate field to matter may result
in local anomalous accelerations of order $\Lambda^2/M\sim H_0$. This
may be an interesting or deadly feature of this model, depending on whether
such effects are present inside the Solar system. The detailed study of these
effects is beyond the scope of this paper.

\section{Summary}\label{summary}

We have studied the connection between the null energy condition and the microscopic stability
of the system that sources the energy momentum tensor. We have not considered extra pathologies that may arise through the gravitational backreaction.
Our attention has been limited to a general system of derivatively coupled scalar fields, but we expect that our arguments
can be generalized to include vector fields. However, we have explained that the class of theories we consider includes relativistic solids and fluids, and is thus relevant to cosmology.

We have first worked under the assumptions that all gapless scalar modes have a non-degenerate dispersion relation $\omega= v k$ with $v\not =0,\infty$.
Our main result then is that a violation of the null energy condition does not strictly imply
instabilities (ghosts or imaginary frequencies) in the system. We have proven that by
producing counter-examples, that is acceptable effective field theories with
a positive definite Hamiltonian for quadratic perturbations around a background whose
energy momentum violates the null energy condition. A necessary feature of all the counter-examples is the anisotropy of the background and, perhaps more importantly, the presence
of superluminal modes. In fact we proved that for systems that are either isotropic 
or do not feature superluminality, a violation of the null energy condition always implies an unescapable instability.
From a purely effective field theoretic viewpoint, in the presence of
a Lorentz violating background, there does not seem to exist any limitation on the presence
of superluminal modes. Indeed  many (most?) of the models that have been
considered in recent years in connection to cosmic acceleration  \cite{Armendariz-Picon:2000ah,Dvali:2000hr,Arkani-Hamed:2003uy} feature superluminal excitations around some background.
Our study suggests that effective field theories that allow superluminality are subtly distinct
from those which do not. The apparent paradoxes that we discussed in connection to causality
are another sign of this distinction and a simple, but deep, characterization of it
will be given in Ref.~\cite{superluminal}.

We have then considered isotropic systems in which some mode may have a
degenerate dispersion relation, either $\omega=0$ or $k=0$. We expect such
systems to be acceptable only when the degenerate dispersion relations are
protected by symmetries. We have studied various internal symmetries of the
scalar field manifold.  These symmetries nicely generalize to relativistic
solids the internal symmetry that distinguishes ordinary liquids from ordinary
solids (see paragraph below).  Although we have not been able to perform a
completely general analysis, we believe we have given fairly convincing
evidence that a violation of the null energy condition leads to instability
even in the presence of such symmetries. It is interesting, in this respect,
that the so called theories of massive gravity, broadly defined by having a
non-trivial field background while the energy momentum tensor vanishes
exactly, lie right at the boundary between theories that violate and theories
that respect the null energy condition. This feature largely explains the
difficulty to make such theories viable.

Aside the above results, two other directions were partially investigated.
In sect.\ref{fluids} we considered the subclass of scalar field theories involving three scalar fields
$\phi^I$, $I=1,2,3$, with space coordinate dependent expectation values $\langle \phi^I(x)\rangle
=x^I$. We explained that these systems are naturally interpreted as solids filling the universe.
We  characterized the limit where the solid becomes a fluid by the appearance of a volume
preserving reparametrization invariance of the scalar field manifold.
 We then generalized
to the relativistic case in a Lagrangian formulation some of the well known results of
fluid mechanics, like Kelvin's theorem. We also discussed a sort of duality relation
between the recently proposed ghost condensate model and an ordinary  diluted fluid. 
In this respect the novelty of the ghost condensate models partially lies in its being
a weakly coupled theory, which is not the case for the corresponding diluted fluid model.
In our opinion, the results in sect. \ref{fluids}, while not completely original and  not fully clear cut, are still
useful as they put in a wider perspective the classes of scalar theories we have been working  on.

Finally in section \ref{towards} we briefly stepped into another class of theories, where one scalar with non-trivial
time dependent background is directly coupled to the matter energy momentum and effectively mimics 
a cosmic acceleration with $\dot H>0$.  We have simply explained the two difficulties that have made past attempts 
of this type very clumsy. These are due to
the tests of general relativity and to the gravitational backreaction of the metric 
on the scalar field energy momentum. The two difficulties point to a simple solution, where the scalar field
is a ghost condensate field, for which the energy momentum can be exactly zero  thus causing no backreaction.
The possible difficulties of such a model lie in the non-linear dynamics of the ghost condensate in the presence of massive bodies, though it has reasonably been argued that such non-linearities may not necessarily lead to trouble.
 Our simple proposal potentially opens the way to genuine modifications of gravity
that make $w<-1$ consistent, but further study is needed.

%%%%%%%%%%%%%%%%%%%%%%%%%%%%%%%%%%%%%%%%%%%%%%%%%%%%%%%%%%%%%%%%%%%%%%%%%%%%%%
%%%%%%%%%%%%%%%%%%%%%%%%%%%%%%%%%%%%%%%%%%%%%%%%%%%%%%%%%%%%%%%%%%%%%%%%%%%%%%

%\section{Concluding remarks}
%\label{conclusions}
\section*{Acknowledgments}
We would like to thank N.~Arkani-Hamed, P.~Creminelli, V.~Rubakov, I.~Tkachev, L.~Senatore,
T.~Wiseman and M.~Zaldarriaga for useful conversations.
This work has been partly supported by the European Commission under contract  MRTN-CT-2004-503369.

\appendix
\section{Stability of the NEC-violating systems of section~\ref{examples}}
In section \ref{examples}, we presented two Lagrangians which were claimed to
be stable despite violating the NEC. Here we sketch the stability analysis of these systems.
\subsection{Two dimensional example}
The Lagrangian for the two dimensional example is given by
\be
\label{2dexample1}
{\cal L}=\left(\tilde B^{00}+ \tilde B^{11}+2(1+a^2)\, \tilde B^{01}-(a^2+b^2)\, \tilde B^{00} \tilde B^{01}+
(a^2+c^2)\, \tilde B^{11} \tilde B^{01}\right)\;.
\ee
with \[
b^2 <2,\qquad c^2<2\;.
\] 
To study the stability of the fluctuations, it is useful to know the expansion of
 the meson field $\tilde{B}^{IJ}$,
\begin{gather}
\label{2dBIJ}
\tilde B^{00}=2 \, \d_0\pi^0 + \partial_\mu \pi^0 \partial^\mu \pi^0 \nonumber\\
\tilde B^{11}=-2 \, \d_1\pi^1+ \partial_\mu \pi^1 \partial^\mu \pi^1 \nonumber\\
\tilde B^{01}=(\d_0\pi^1-\d_1\pi^0) + \partial_\mu \pi^0\partial^\mu \pi^1        \nonumber
\end{gather}
The Lagrangian for the fluctuations can then be written as
\begin{multline}
\label{2dexamplefluc}
\l\d_\mu\pi^0\r^2+\l\d_\mu\pi^1\r^2+2(1+a^2)\l\d_\mu\pi^0\r\l
\d^\mu\pi^1\r \\
-(a^2+b^2) 2 \partial_0 \pi^0 \partial_0 \pi^1 +(a^2+c^2) \partial_1 \pi^1 \partial_1 \pi^0 
 + (a^2+c^2) 2 \partial_0 \pi^0 \partial_1 \pi^0 - (a^2+c^2) \partial_1 \pi^1 \partial_0 \pi^1 \;.
\end{multline}
The first line of \eqref{2dexamplefluc} comes from the first three terms of
\eqref{2dexample1} which are linear in $\tilde{B}^{IJ}$. Those are the only
term of \eqref{2dexample1} that contribute to the background energy-momentum
tensor and are responsible for the NEC violation. We can see that taken
alone, they would form an unstable system as both the kinetic and gradient
matrices of the first line of \eqref{2dexamplefluc} have negative eigenvalues,
due to the mixing between $\pi^0$ and $\pi^1$. But this defect is cured by the
terms $\tilde{B}^{00}\tilde{B}^{01} $ and $\tilde{B}^{11}\tilde{B}^{01} $ that
provide terms in the fluctuation Lagrangian that suppress the kinetic and
gradient mixing respectively. They also give terms with one space and one time
derivative which do not affect the positivity of the Hamiltonian
(cf. eq.~(\ref{hamiltonian})).
 
\subsection{Three dimensional example}
The Lagrangian written in terms of $\tilde{B}^{IJ}$ has form
\be
\label{finalaction1}
{\cal L} = T_{IJ} \, \tilde B^{IJ}+ \left( \alpha^2 \big[(\tilde B^{01})^2-(\tilde
B^{11})^2-(\tilde B^{12})^2 \big] +ch\, sh \, (1+a^2) \big[\tilde
B^{02}(-\tilde B^{11}+\tilde B^{22})+ 2\tilde B^{01}\tilde B^{12} \big] \right)\; 
\ee
with $a^2\sim b^2\sim\alpha^2\sim 1$ and $T_{IJ}$ as given by
eq.~(\ref{boostedT}).  As in the two dimensional example, terms linear in
$\tilde{B}^{IJ}$ give a Lagrangian for the fluctuation of the form $T_{IJ}
\partial_\mu\pi^I \partial^\mu \pi^J$ that is Lorentz invariant and unstable,
due to the $-b^2 (\partial_\mu \pi^1)^2$ that it contains. And as before, this
can be cured by adding the terms that are quadratic in $\tilde{B}^{IJ}$. To
understand what these terms do, it is useful to keep in mind the expansions of
the $\tilde{B}^{IJ}$ to linear order in $\pi^I$,
\begin{gather}
 \tilde  B^{00}=2 \, \d_0\pi^0\nonumber\\
 \tilde B^{11}=-2 \, \d_1\pi^1\nonumber\\
 \tilde B^{22}=-2 \, \d_2\pi^2\nonumber\\
 \tilde B^{01}=\d_0\pi^1-\d_1\pi^0\nonumber\\
 \tilde B^{12}=-\d_1\pi^2-\d_2\pi^1\nonumber\\
 \tilde B^{02}=\d_0\pi^2-\d_2\pi^0\nonumber
\end{gather}

The kinetic part of the Lagrangian for the fluctuations is given by
\begin{equation}
K = (ch^2 a^2 + sh^2) \left(\partial_0 \pi^0 \right)^2 + 
2 ch sh (1+a^2) \partial_0 \pi^0 \partial_0 \pi^2 + 
(ch^2+ a^2 sh^2) \left(\partial_0 \pi^2 \right)^2 +(\alpha^2 - b^2) \left(\partial_0 \pi^1 \right)^2 \;,
\end{equation}
while the gradient matrix is
\begin{multline}
\label{G}
G=(ch^2a^2+sh^2) \, (\d_i\pi^0)^2+(sh^2a^2+ch^2) \, (\d_i\pi^2)^2+
(\alpha^2-b^2) \, (\d_i\pi^1)^2 \\
- \alpha^2 \left[\left(\partial_1 \pi^0 \right)^2 + \partial_1 \pi^1 \partial_2 \pi^2 - \left(\partial_1 \pi^2 \right)^2 \right]
\end{multline}

The role of the term $\alpha^2(\tilde{B}^{01})^2$ in \eqref{finalaction1} is 
to make the kinetic matrix positive definite, which is the case if 
\be
\label{alpha}
\alpha^2-b^2>0\;.  
\ee The term $ ch\, sh \, (1+a^2) \big[\tilde
B^{02}(-\tilde B^{11}+\tilde B^{22})+ 2\tilde B^{01}\tilde B^{12} \big]$
removes mixing between $\pi^2$ and $\pi^0$ in the gradient matrix. This is
needed because with the mixing the gradient matrix, despite having entries of
order $ch \, sh$ , has an eigenvalue that is small and the term $- \alpha^2
(\partial_1 \pi^0)^2$ contained in $\alpha^2 (\tilde{B}^{01} )^2$ could make
this matrix non-positive definite. Note, that mixing between  $\pi^2$ and $\pi^0$ is
still present in the kinetic matrix. As a result this matrix has a small eigenvalue leading
to the highly superluminal mode in the spectrum in agreement with our results.

Finally, the terms $-\alpha^2 \left[(\tilde B^{11})^2-(\tilde
B^{12})^2\right]$ are there to change the sign of the gradient term for
$(\partial_i \pi^1)^2$.  The terms on the last line of \eqref{G} are of order
$1$ and subdominant compared to the terms of order $ch, sh$ and do not affect
the positivity of the gradient matrix.


\begin{thebibliography}{99}
\bibitem{Hawkingtheorems}
	S.~W.~Hawking and G.~F.~R.~Ellis, 
	{\em The Large Scale Structure of Space-Time},
	Cambridge 1973,
	Cambridge University Press.
\bibitem{Bousso}
  R.~Bousso,
  %``The holographic principle,''
  Rev.\ Mod.\ Phys.\  {\bf 74}, 825 (2002)
  [arXiv:hep-th/0203101].
  %%CITATION = HEP-TH 0203101;%%
\bibitem{Hsu:2004vr}
  S.~D.~H.~Hsu, A.~Jenkins and M.~B.~Wise,
  %``Gradient instability for w<-1,''
  Phys.\ Lett.\ B {\bf 597}, 270 (2004)
  [arXiv:astro-ph/0406043].
  %%CITATION = ASTRO-PH 0406043;%%
  %\cite{Dvali:2000hr}
\bibitem{wormholes} 
  M.~S.~Morris and K.~S.~Thorne,
  %``Wormholes In Space-Time And Their Use For Interstellar Travel: A Tool For
  %Teaching General Relativity,''
  Am.\ J.\ Phys.\  {\bf 56}, 395 (1988).
  %%CITATION = AJPIA,56,395;%%
\bibitem{omega<-1}
  A.~G.~Riess {\it et al.}  [Supernova Search Team Collaboration],
  %``Type Ia Supernova Discoveries at z>1 From the Hubble Space Telescope:
  %Evidence for Past Deceleration and Constraints on Dark Energy Evolution,''
  Astrophys.\ J.\  {\bf 607}, 665 (2004)
  [arXiv:astro-ph/0402512].
  %%CITATION = ASTRO-PH 0402512;%%omega<-1
\bibitem{Dvali:2000hr}
  G.~R.~Dvali, G.~Gabadadze and M.~Porrati,
  %``4D gravity on a brane in 5D Minkowski space,''
  Phys.\ Lett.\ B {\bf 485}, 208 (2000)
  [arXiv:hep-th/0005016].
  %%CITATION = HEP-TH 0005016;%%
%\cite{Sahni:2002dx}
\bibitem{Sahni:2002dx}
  V.~Sahni and Y.~Shtanov,
  %``Braneworld models of dark energy,''
  JCAP {\bf 0311}, 014 (2003)
  [arXiv:astro-ph/0202346].
  %%CITATION = ASTRO-PH 0202346;%%
  %\cite{Lue:2004za}
\bibitem{Lue:2004za}
  A.~Lue and G.~D.~Starkman,
  %``How a brane cosmological constant can trick us into thinking that W <
  %-1,''
  Phys.\ Rev.\ D {\bf 70}, 101501 (2004)
  [arXiv:astro-ph/0408246].
  %%CITATION = ASTRO-PH 0408246;%%
 \bibitem{Nik} A. Nicolis, R. Rattazzi, unpublished.
\bibitem{Arkani-Hamed:2002sp}
N.~Arkani-Hamed, H.~Georgi and M.~D.~Schwartz,
%``Effective field theory for massive gravitons and gravity in theory space,''
Annals Phys.\  {\bf 305}, 96 (2003)
[arXiv:hep-th/0210184].
%%CITATION = HEP-TH 0210184;%%
\bibitem{Arkani-Hamed:2003uy}
N.~Arkani-Hamed, H.~C.~Cheng, M.~A.~Luty and S.~Mukohyama,
%``Ghost condensation and a consistent infrared modification of gravity,''
JHEP {\bf 0405}, 074 (2004)
[arXiv:hep-th/0312099].
%%CITATION = HEP-TH 0312099;%%
%\cite{Rubakov:2004eb}
\bibitem{Rubakov:2004eb}
V.~Rubakov,
%``Lorentz-violating graviton masses: Getting around ghosts, low strong
%coupling scale and VDVZ discontinuity,''
arXiv:hep-th/0407104.
%%CITATION = HEP-TH 0407104;%%
\bibitem{Dubovsky:2004sg}
S.~L.~Dubovsky,
%``Phases of massive gravity,''
JHEP {\bf 0410}, 076 (2004)
[arXiv:hep-th/0409124].
%%\cite{Buniy:2005vh}
\bibitem{Buniy:2005vh}
  R.~V.~Buniy and S.~D.~H.~Hsu,
  %``Instabilities and the null energy condition,''
  arXiv:hep-th/0502203.
  %%CITATION = HEP-TH 0502203;%%%CITATION = HEP-TH 0409124;%%
\bibitem{Nicolis:2004qq}
A.~Nicolis and R.~Rattazzi,
%``Classical and quantum consistency of the DGP model,''
JHEP {\bf 0406}, 059 (2004)
[arXiv:hep-th/0404159].
%%CITATION = HEP-TH 0404159;%%
\bibitem{Dubovsky:2005dw}
  S.~L.~Dubovsky, P.~G.~Tinyakov and I.~I.~Tkachev,
  %``Cosmological attractors in massive gravity,''
  Phys.\ Rev.\ D {\bf 72}, 084011 (2005)
  [arXiv:hep-th/0504067].
  %%CITATION = HEP-TH 0504067;%%
  %\cite{Cline:2003gs}
\bibitem{Cline:2003gs}
  J.~M.~Cline, S.~y.~Jeon and G.~D.~Moore,
  %``The phantom menaced: Constraints on low-energy effective ghosts,''
  Phys.\ Rev.\ D {\bf 70}, 043543 (2004)
  [arXiv:hep-ph/0311312].
  %%CITATION = HEP-PH 0311312;%%
  %\cite{Kaplan:2005rr}
\bibitem{Kaplan:2005rr}
  D.~E.~Kaplan and R.~Sundrum,
  %``A symmetry for the cosmological constant,''
  arXiv:hep-th/0505265.
  %%CITATION = HEP-TH 0505265;%%
%\cite{Arkani-Hamed:2002fu}
\bibitem{Arkani-Hamed:2002fu}
  N.~Arkani-Hamed, S.~Dimopoulos, G.~Dvali and G.~Gabadadze,
  %``Non-local modification of gravity and the cosmological constant problem,''
  arXiv:hep-th/0209227.
  %%CITATION = HEP-TH 0209227;%%
  \bibitem{Wald} R. Wald, {\it General relativity}, The University of Chicago Press, 1984.
\bibitem{jackiw}
  R.~Jackiw, V.~P.~Nair, S.~Y.~Pi and A.~P.~Polychronakos,
  %``Perfect fluid theory and its extensions,''
  J.\ Phys.\ A {\bf 37}, R327 (2004)
  [arXiv:hep-ph/0407101].
  %%CITATION = HEP-PH 0407101;%%
\bibitem{son}
  D.~T.~Son,
  %``Effective Lagrangian and Topological Interactions in Supersolids,''
  Phys.\ Rev.\ Lett.\  {\bf 94}, 175301 (2005)
  [arXiv:cond-mat/0501658].
  %%CITATION = COND-MAT 0501658;%%
\bibitem{weinberg}
S.~Weinberg, {\it Gravitation and Cosmology}, John Wiley \& Sons, 1972.

\bibitem{LL} 
 L.~D.~Landau, E.~M.~Lifshitz,
  {\it Textbook On Theoretical Physics. Vol. 6: Fluid Mechanics}.
  
\bibitem{Son:2002zn}
  D.~T.~Son,
  %``Low-energy quantum effective action for relativistic superfluids,''
  arXiv:hep-ph/0204199.
  %%CITATION = HEP-PH 0204199;%%
%\cite{Krotov:2004if}
\bibitem{Krotov:2004if}
  D.~Krotov, C.~Rebbi, V.~A.~Rubakov and V.~Zakharov,
  %``Holes in the ghost condensate,''
  Phys.\ Rev.\ D {\bf 71}, 045014 (2005)
  [arXiv:hep-ph/0407081].
  %%CITATION = HEP-PH 0407081;%%
%\cite{Arkani-Hamed:2005gu}
\bibitem{Arkani-Hamed:2005gu}
  N.~Arkani-Hamed, H.~C.~Cheng, M.~A.~Luty, S.~Mukohyama and T.~Wiseman,
  %``Dynamics of gravity in a Higgs phase,''
  arXiv:hep-ph/0507120.
  %%CITATION = HEP-PH 0507120;%%
\bibitem{Dubovsky:2004ud}
  S.~L.~Dubovsky, P.~G.~Tinyakov and I.~I.~Tkachev,
  %``Massive graviton as a testable cold dark matter candidate,''
  Phys.\ Rev.\ Lett.\  {\bf 94}, 181102 (2005)
  [arXiv:hep-th/0411158].
  %%CITATION = HEP-TH 0411158;%%
%\cite{Csaki:2005vq}
\bibitem{Csaki:2005vq}
  C.~Csaki, N.~Kaloper and J.~Terning,
  %``The accelerated acceleration of the universe,''
  arXiv:astro-ph/0507148.
  %%CITATION = ASTRO-PH 0507148;%%


\bibitem{Carroll:2003st}
  S.~M.~Carroll, M.~Hoffman and M.~Trodden,
  %``Can the dark energy equation-of-state parameter w be less than -1?,''
  Phys.\ Rev.\ D {\bf 68} (2003) 023509
  [arXiv:astro-ph/0301273].
  %%CITATION = ASTRO-PH 0301273;%%
  %\cite{Khoury:2003rn}
  
  %\cite{Boisseau:2000pr}
\bibitem{Boisseau:2000pr}
  B.~Boisseau, G.~Esposito-Farese, D.~Polarski and A.~A.~Starobinsky,
  %``Reconstruction of a scalar-tensor theory of gravity in an accelerating
  %universe,''
  Phys.\ Rev.\ Lett.\  {\bf 85}, 2236 (2000)
  [arXiv:gr-qc/0001066].
  %%CITATION = GR-QC 0001066;%%
  %%Cited 78 times in SPIRES-HEP

  
  
  
  
\bibitem{Khoury:2003rn}
  J.~Khoury and A.~Weltman,
  %``Chameleon cosmology,''
  Phys.\ Rev.\ D {\bf 69}, 044026 (2004)
  [arXiv:astro-ph/0309411].
  %%CITATION = ASTRO-PH 0309411;%%
  \bibitem{Armendariz-Picon:2000ah}
  C.~Armendariz-Picon, V.~Mukhanov and P.~J.~Steinhardt,
  %``Essentials of k-essence,''
  Phys.\ Rev.\ D {\bf 63}, 103510 (2001)
  [arXiv:astro-ph/0006373].
  %%CITATION = ASTRO-PH 0006373;%%
  \bibitem{superluminal} A. Adams, N. Arkani-Hamed, S. Dubovsky, A. Nicolis, R. Rattazzi,
  paper in preparation.
\end{thebibliography}
\end{document}